\documentclass[twocolumn]{aastex631}

\graphicspath{{./}{figures/}}

\newcommand\dbb{{\texttt{DISKBB}}}
\newcommand\thc{{\texttt{THCOMP}}}

\newcommand\redd{{\texttt{REDDEN}}}
\newcommand\iref{{\texttt{IREFLECT}}}

\newcommand\tba{{\texttt{TBABS}}}
\newcommand\xabs{{\texttt{XABS}}}
\newcommand\tbp{{\texttt{TBPCF}}}
\newcommand\zpo{{\texttt{ZPOWERLAW}}}
\newcommand\zbb{{\texttt{ZBLACKBODY}}}
\newcommand\con{{\texttt{CONSTANT}}}
\newcommand\kd{{\texttt{KERRD}}}

\newcommand\hst{{\it HST}}
\newcommand\chandra{{\it Chandra}}

\newcommand\asat{{\it AstroSat}}

\newcommand\s{{\rm~s}}

\newcommand\kms{{$\rm~km~s^{-1}$}}
\newcommand\ergA{{$\rm erg~cm^{-2}~s^{-1}$\AA$^{-1}$}}
\newcommand\ergS{{$\rm erg~cm^{-2}~s^{-1}$}}

\newcommand\kev{{\rm~keV}}

\newcommand\angs{{\rm~\AA}}
\newcommand\msun{{\rm~M_{\odot}}}
\usepackage{booktabs, makecell}
\usepackage{booktabs}
\usepackage{array}
\usepackage{verbatim}
\usepackage{amsmath}
\usepackage{multirow}
\usepackage{array}
\usepackage{enumitem}
\usepackage{hyperref}
\newcolumntype{H}{>{\setbox0=\hbox\bgroup}c<{\egroup}@{}}
\usepackage{blindtext}
\usepackage{soul}
\usepackage{cancel}

\begin{document}

\title{Multi-epoch UV -- X-ray spectral study of NGC~4151 with \asat{}}


\author{Shrabani Kumar}
\affiliation{Inter-University Centre for Astronomy and Astrophysics (IUCAA), PB No.4, Ganeshkhind, Pune-411007, India}

\author[0000-0003-1589-2075]{G. C. Dewangan}
\affiliation{Inter-University Centre for Astronomy and Astrophysics (IUCAA), PB No.4, Ganeshkhind, Pune-411007, India}

\author[0000-0003-3105-2615]{P. Gandhi}
\affiliation{School of Physics and Astronomy, University of Southampton, Highfield, Southampton SO17 1BJ, UK}

\author{I. E. Papadakis}
\affiliation{Department of Physics and Institute of Theoretical and Computational Physics, University of Crete, 71003 Heraklion, Greece}
\affiliation{Institute of Astrophysics, FORTH, GR-71110 Heraklion, Greece}

\author{N. P. S. Mithun}
\affiliation{Physical Research Laboratory Thaltej, Ahmedabad, Gujarat 380009, India}

\author{K. P. Singh}
\affiliation{Indian Institute of Science Education and Research Mohali, Knowledge City, Sector 81, Manauli P.O., SAS Nagar, 140306, Punjab, India}
\affiliation{Department of Astronomy and Astrophysics, Tata Institute of Fundamental Research, 1 Homi Bhabha Road, Mumbai 400005, India}

\author{D. Bhattacharya}
\affiliation{Ashoka University, Dept. Of Physics, Sonipat, Haryana-131029, India}

\author[0000-0002-0333-2452]{A. A. Zdziarski}
\affiliation{Nicolaus Copernicus Astronomical Center, Polish Academy of Sciences, Bartycka 18, PL-00-716 Warszawa, Poland}

\author[0000-0001-9097-6573]{G. C. Stewart}
\affiliation{Department of Physics and Astronomy, The University of Leicester, University Road, Leicester LE1 7RH, UK}

\author{S. Bhattacharyya}
\affiliation{Department of Astronomy and Astrophysics, Tata Institute of Fundamental Research, 1 Homi Bhabha Road, Mumbai 400005, India}

\author{S. Chandra}
\affiliation{Center for Space Research, North-West University, Potchefstroom 2520, South Africa}

\begin{abstract}
We present a multi-wavelength spectral study of NGC 4151 based on five epochs of simultaneous \asat{} observations in the near ultra-violet (NUV) to hard X-ray band ($\sim 0.005-80$ keV) during $2017 - 2018$. We derived the intrinsic accretion disk continuum after correcting for internal and Galactic extinction, contributions from broad and narrow line regions, and emission from the host galaxy. We found a bluer continuum at brighter UV flux possibly due to variations in the accretion disk continuum or the UV reddening. We estimated the intrinsic reddening, $E(B-V) \sim 0.4$, using high-resolution \hst{}/STIS spectrum acquired in March 2000. We used thermal Comptonization, neutral and ionized absorption, and X-ray reflection to model the X-ray spectra. We obtained the X-ray absorbing neutral column varying between $N_H \sim 1.2-3.4 \times 10^{23} cm^{-2}$, which are $\sim 100$ times larger than that estimated from UV extinction, assuming the Galactic dust-to-gas ratio. To reconcile this discrepancy, we propose two plausible configurations of the obscurer: (a) a two-zone obscurer consisting of dust-free and dusty regions, divided by the sublimation radius, or (b) a two-phase obscurer consisting of clumpy, dense clouds embedded in a low-density medium, resulting in a scenario where a few dense clouds obscure the compact X-ray source substantially, while the bulk of UV emission arising from the extended accretion disk passes through the low-density medium. Furthermore, we find a positive correlation between X-ray absorption column and $NUV-FUV$ color and UV flux, indicative of enhanced winds possibly driven by the `bluer-when-brighter' UV continuum.



  

\end{abstract}


\section{Introduction} \label{sec:intro}
NGC~4151, a type 1.5 Seyfert galaxy, is one of the brightest and nearest (z $\sim 0.0033$, distance  $\sim 15.8$ Mpc) active galactic nuclei (AGN). It has been studied quite extensively in different wavebands \citep{perola1986new,zdziarski1996origin,ulrich2000active,zoghbi2019revisiting,mahmoud2020discarding}. The X-ray spectrum shows complex absorption due to multiple layers of ionized and neutral absorbing columns  \citep{zdziarski1996origin,puccetti2007rapid,zoghbi2019revisiting,kraemer2020mass}. The shape of the X-ray spectrum, particularly in the $1-6$~keV band \citep{beuchert2017suzaku}, is observed to be significantly modified mostly due to the variation in the neutral absorption column density ($N_H \sim 10^{22}-10^{23}~\rm cm^{-2}$). Large changes in the column density (by a factor of $\sim 10$) are observed to happen on timescales of days to months \citep{puccetti2007rapid}. The soft X-ray excess emission, observed below 2 keV, is nearly constant in flux \citep{1996ApJ...470..364E,zoghbi2019revisiting}. This emission component, spatially resolved by Chandra, is observed to be arising from a few hundred-parsec distance off an extended region \citep{yang2001chandra}. The X-ray photon index is observed to vary between $1.4-1.8$ \citep{zoghbi2019revisiting}. The X-ray reflection spectrum shows the ubiquitous presence of a narrow Fe K$\alpha$ line at 6.4~keV, implying distant reflection (possibly due to the torus) of the primary X-ray continuum. 
\citet{2010MNRAS.408.1851L} observed the relative reflection strength increasing at lower flux. They attributed the variation in the reflection strength to the nearby reflection from the disk.  In the bright flux state, weak relativistic reflection is observed, which could originate from the disk truncated around 10$r_g$ \citep{szanecki2021relativistic}. The distant reflection component is observed to be constant with a reflection fraction of $\sim 0.3$ \citep{2010MNRAS.408.1851L}.

The UV/optical spectrum shows multiple broad and narrow emission and absorption lines. The broad emission lines (\ion{C}{4}, \ion{Mg}{2}, H$\alpha$, H$\beta$) are associated with appearing and disappearing, blue and red shifted, wings, which also vary in strength over time \citep{ulrich1996month,metzroth2006mass,shapovalova2008long}.  This indicates a change in the BLR kinematics in response to continuum strength \citep{ulrich1991ultraviolet,ulrich2000active,bon2012first,chen2023broad}. The continuum flux shows an order of magnitude variability on a timescale of a few years. 
The optical continuum flux (at 5100 \AA) since 1993 has shown two minima, in $2000-2001$ and $2005-2006$ \citep{chen2023broad}. The maximum UV/Optical flux was observed during $1995-1996$, and the other high flux states were observed during 2003 and 2010. From 1996 to 2006, the continuum and line flux varied by a factor of $\sim 6$. The historical peak continuum flux (in UV) observed during 1995 is $\sim 5 \times 10^{-13}$ \ergA, and the minimum flux observed in $2000-2001$ was $\sim 10^{-14} $ \ergA. During this flux change, the spectral state changed from  Seyfert 1.5 (maximum flux) to 1.8 (minimum flux), the reason for which could be attributed to the change in BLR radiation pressure \citep{chen2023broad}.   

Flux variation has similarly been observed in the X-ray emission. \citet{2010MNRAS.408.1851L} categorized the source into three states: bright, medium, and dim, according to the level of X-ray flux. The X-ray flux was at its maximum (bright state) around 1993, 2003, and 2009 and at its minimum in 2000, 2006, and 2007 (dim state). The maximum X-ray flux observed during 1993 was $\sim 10^{-10}$ \ergS~ in the $2-10$ keV band. The electron temperature of the X-ray emitting hot corona is $\sim 50-70$ keV in the bright state and $\sim 180-230$ keV in the dim state \citep{2010MNRAS.408.1851L}. 

The UV/optical and X-ray variations followed similar trends. The UV/optical and X-ray flux correlate in simultaneous UV/X-ray observations \citep{perola1986new,2010MNRAS.408.1851L}. 
Here, we investigate the connection between UV and X-ray spectral variability. 
 We analyzed the five sets of UV/X-ray data simultaneously acquired by \asat~ during 2017 -- 2018. We studied the effect of X-ray absorbing material on UV emission and the connection between the UV and X-ray emission. 
  We describe the observation and data processing in Section~\ref{sec:obs}, we present the spectral analysis in Section~\ref{sec:uvitspec}, \ref{sec:xrayspec} and \ref{sec:uvXrayspec}, and describe our results and discuss them in Section~\ref{sec:res_disc}. 

\begin{deluxetable*}{cccccccccc}
\tablenum{1}
\tablecaption{Log of \asat{} observations of NGC~4151. The energy bands used to calculate the net count rates are $7-9.5$ eV (UVIT/Grating), $0.7-7$ keV  (SXT), $4-20$ keV (LAXPC), and $22-80$ keV (CZTI).}
\label{tab:astrosatlog}
\tablewidth{700pt}
\tablehead{
Obs ID & Date of Obs & Instrument & Count Rate & Exposure Time\\
\multicolumn2c{}&  &$\rm (counts~s^{-1})$&(ks)
}
\startdata
G06\_117T01\_9000001012 & 07-02-2017& FUV-BaF$_2$ & $37.6\pm0.2$ & 0.9\\
(obs1)&&NUV-B15 & $10.2\pm 0.02$ & 22.7\\
&&SXT & $0.238\pm0.004$& 22.7 \\
&&LAXPC & $21.85\pm0.05$& 30.7\\
&&CZTI& $1.0\pm 0.1$& 23.3\\
G06\_117T01\_9000001046 & 22-02-2017& FUV-BaF$_2$ & $31.44\pm0.04$ & 23.6\\
(obs2)&&NUV-B15 & $8.7\pm 0.02$ & 23.2\\
&&SXT & $0.211\pm0.003$&31.9\\
&&LAXPC& $20.31\pm0.05$&50\\
&&CZTI&$0.85\pm0.08$&54.7\\
G06\_117T01\_9000001086 &16-03-2017& FUV-BaF$_2$& $26.19\pm0.03$ & 23.9\\
(obs3)&&NUV-B15& $7.39\pm0.02$&15.9\\
&&SXT & $0.181\pm0.003$ & 31.6\\
&&LAXPC & $14.12\pm 0.04$& 47.2\\
&&CZTI & $0.57\pm0.08$ & 53\\
G08\_064T01\_9000001814 & 03-01-2018&FUV-G1& $5.72\pm0.04$ & 3.8\\
(obs4)&&FUV-G2& $9.14\pm0.05$ &3.7 \\
&&FUV-BaF$_2$& $21.32\pm0.05$& 8.9\\
&&NUV-B15& $6.67\pm0.02$&14.8 \\
&&SXT & $0.239\pm0.003$ & 28.5\\
&&LAXPC & $17.71\pm0.04$&59.2\\
&&CZTI & $0.75\pm0.02$&74.9\\
G08\_064T01\_9000002070 & 02-05-2018&FUV-G1& $6.82\pm0.04$ &3.9\\
(obs5)&&FUV-G2& $11.13\pm0.05$&3.4\\
&&FUV-BaF$_2$& $25.88\pm0.02$&9.4\\
&&SXT& $0.239\pm0.003$&30\\
&&LAXPC& $18.11\pm0.08$&54.8\\
&&CZTI & $0.51\pm0.08$ & 44.3
\enddata
\end{deluxetable*}

\section{\asat{}~ Observations and Data reduction} \label{sec:obs}

 \asat{} \citep{singh2014astrosat} is a multi-wavelength space observatory launched in 2015 by the Indian Space Research Organization (ISRO). It carries four co-aligned payloads, namely, Cadmium-Zinc-Telluride Imager (CZTI; \citealt{czti,bhalerao2017cadmium}), Large Area X-ray Proportional Counter (LAXPC; \citealt{yadav2016large,antia2017calibration}), Soft X-ray Telescope (SXT; \citealt{singh2016orbit,singh2017soft}), and Ultra-violet Imaging Telescope (UVIT; \citealt{tandon2017orbit,tandon2020additional}). \asat{} observed NGC~4151 five times during February 2017 -- May 2018 with all four co-aligned payloads. We list the details of the five sets of simultaneous UV/X-ray observations G06\_117T01\_9000001012 (obs1), G06\_117T01\_9000001046 (obs2), G06\_117T01\_9000001086 (obs3), G08\_064T01\_9000001814 (obs4) and G08\_064T01\_9000002070 (obs5) in Table~\ref{tab:astrosatlog}. 

 \begin{figure}[ht!]
 \epsscale{1.1}
\plotone{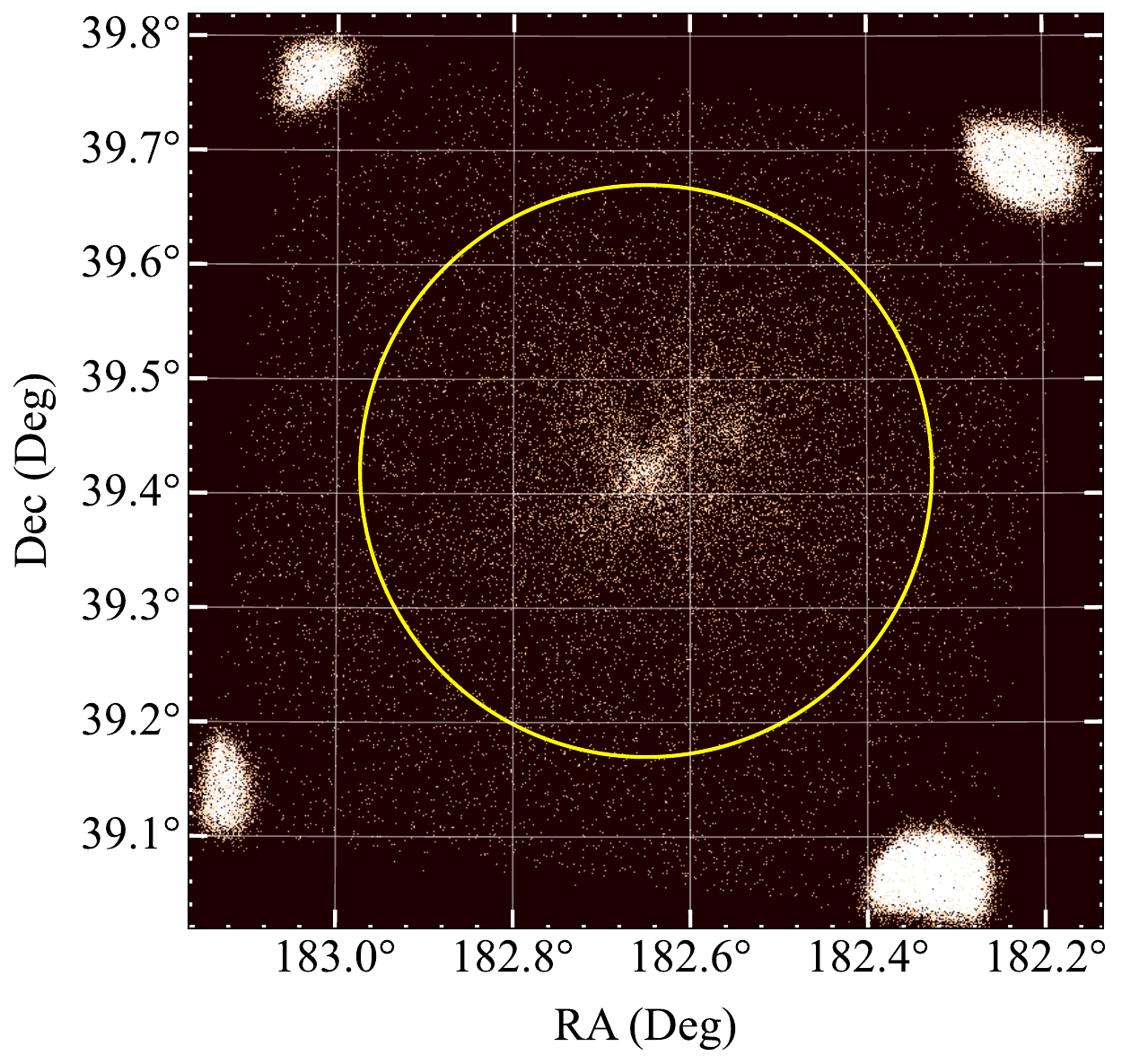}
\caption{The SXT image of NGC~4151 (obs4), the yellow circle (radius $= 15'$) shows the source extraction area.
\label{fig:sxt}}
\end{figure}
 
\subsection{SXT}
\label{subsec:sxtdata}
The SXT is a focusing soft  X-ray telescope with a CCD similar in design to the XRT onboard the {\it Swift} observatory. It observes mainly in photon counting (PC) mode in the $0.3-7$ keV band. The field of view is $\sim 40'$, the half power diameter is $\sim 11'$, and the energy resolution is 150 eV at $6$ keV \citep{singh2017soft}. We used level1 data and generated the level2 clean event files using SXT pipeline software AS1SXTLevel2-1.4b available at the SXT payload operation center (POC\footnote{\url{https://www.tifr.res.in/~astrosat\_sxt/sxtpipeline.html}}). For each observation, we merged the clean event files from every orbit using the SXT merger tool  SXTMerger\footnote{\url{https://github.com/gulabd/SXTMerger.jl}}, also available at the SXT POC website. The final cleaned image of the source corresponding to obs4 is shown in Figure~\ref{fig:sxt}. We used the XSELECT tool within HEASoft (version 6.29) to extract the source spectrum from the circular region of $15'$ radius. We used the background spectrum (SkyBkg\_comb\_EL3p5\_Cl\_Rd16p0\_v01.pha), instrument response (RMF: sxt\_pc\_mat\_g0to12.rmf), and effective area (ARF: sxt\_pc\_excl00v04\_20190608.arf) from the SXT POC website. We grouped each  PHA spectral dataset with a minimum $25~\rm counts~bin^{-1}$ using the tool FTGROUPPHA available under the HEASoft package. The net source count rate varies from $0.18-0.24~\rm counts~s^{-1}$ in the $0.7-7$ keV energy band (Table~\ref{tab:astrosatlog}).

\subsection{LAXPC}
\label{subsec:laxpcdata}

The LAXPC consists of three gas-proportional counters (LAXPC10, LAXPC20, and LAXPC30) and is sensitive to the $3-80$ keV energy band. We used the data acquired by LAXPC20 as the data from the other two instruments are not usable since LAXPC10 is unstable due to continuous variation of gain, and the LAXPC30 has significant gas leakage in the detector \citep{antia2017calibration}.  We used the pipeline software LAXPCSOFT V3.4.4 \footnote{\url{https://www.tifr.res.in/~astrosat\_laxpc/LaxpcSoft.html}} to generate the spectrum and a suitable background. We used the response file  lx20cshm08L1v1.0.rmf for our spectral analysis. We grouped the PHA spectral data to a minimum of $20~\rm counts~bin^{-1}$ using the tool FTGROUPPHA available in the HEASoft package. The net source count rate varied between $14-22~\rm counts~s^{-1}$ in the $4-20$ keV band during our observations (see Table~\ref{tab:astrosatlog}).

\subsection{CZTI}
\label{subsec:cztidata}
The CZTI is a hard X-ray telescope that uses coded mask imaging. The imager consists of four quadrants with 16 CZT detector modules in each of them. It operates in photon counting mode and observes in the $20-100$ keV energy band. The field of view is $4.6^\circ \times 4.6^\circ$ and the energy resolution is $8\%$ at 100 keV. We processed the CZTI data using the
 pipeline version 3.0 along with the associated CALDB. Following the standard pipeline procedure, we obtained the event files, which we used to generate background-subtracted source spectra for each quadrant (along with associated
response matrices)  by employing the
mask-weighting technique.

\subsection{UVIT}
\label{subsec:uvitdata}
The UVIT comprises three channels: far ultraviolet (FUV: $1200-1800$\angs), near ultraviolet (NUV: $2000-3000$\angs), and visible (VIS: $3200-5500$\angs). Both the FUV and NUV channels are equipped with several broadband filters that provide high-resolution (FWHM $\sim 1 -1.5$ arcsec) images in $28$ arcmin diameter field \citep{tandon2017orbit,tandon2020additional}. The FUV channel also includes two low spectral resolution slit-less gratings (hereafter FUV-G1 and FUV-G2) oriented orthogonal to each other; the NUV channel has one grating (NUV-G). The calibration of these gratings is described in \cite{Dewangan_2021}, and the analysis tools are included in the UVITTools.jl package\footnote{\url{https://github.com/gulabd/UVITTools.jl}} with updated calibration. The spectral resolution of the FUV gratings in the $-2$ order is $\sim 14.3$\angs{}. We used the available broadband filter observation F154W (FUV-BaF$_2$, $\lambda_m = 1541$\angs, $\Delta \lambda = 380$\angs) in the FUV channel for all the five observations, and N219M (NUV-B15, $\lambda_m= 2197$\angs, $\Delta \lambda= 270$\angs) in the NUV channel for four of our observations (excluding the obs5), since the NUV detector stopped functioning in 2018, March \citep{ghosh2021orbit}. Additionally,  we used FUV grating observations, which were available for obs4 and obs5 only.

\begin{figure}[ht!]
 \epsscale{1.1}
\plotone{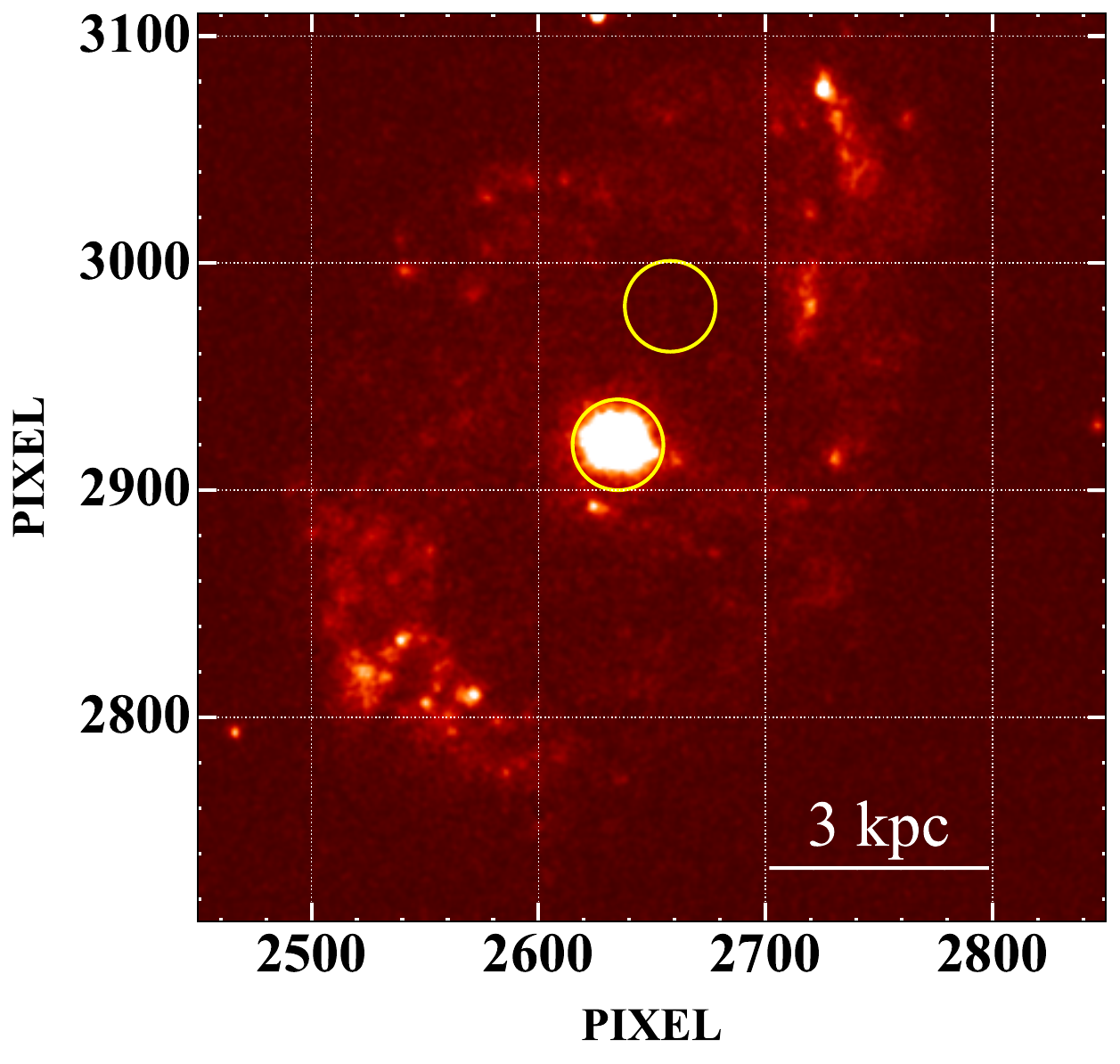}
\caption{NUV-B15 image of NGC~4151, the yellow circles correspond to a radius of 20 pixels ($8.3^{\prime \prime}$; 1 pixel $\sim 0.416^{\prime\prime}$). The background is taken from within the galaxy with a relatively clean region.   
\label{fig:nuv}}
\end{figure}

We processed the level1 UVIT data using the CCDLAB pipeline software \citep{Postma_2017}. This pipeline generates a single image for each orbit, implements rotation and translation to co-align the orbit-wise images, considering one of the images as a reference, and then merges into one final image. We obtained the final images for both the broadband filters and gratings.  We extracted the photometric flux for the broadband filter data using the UVITTools.jl package after correcting for the saturation effect on the source following \citet{tandon2020additional} for all five observations. We calculated the flux within a circular region of radius  $8.3^{\prime\prime}$  centered at the source centroid position. We obtained the background flux by using an identical circular area from within the host galaxy (see Figure~\ref{fig:nuv}). We purposefully used the background region from within the galaxy as the diffuse emission from the galaxy may have contaminated the AGN emission as well. For obs1, CCDLAB could not generate the final image in the FUV filter. Therefore, we used one of the orbit images with the longest exposure to calculate the observed flux and generated the single-channel spectrum. In Figure~\ref{fig:nuv}, we show the NUV broadband filter image from one of the observations where yellow circles mark the source and background extraction regions used to obtain the photometric flux. 

\begin{figure}[ht!]
 \epsscale{1.1}
\plotone{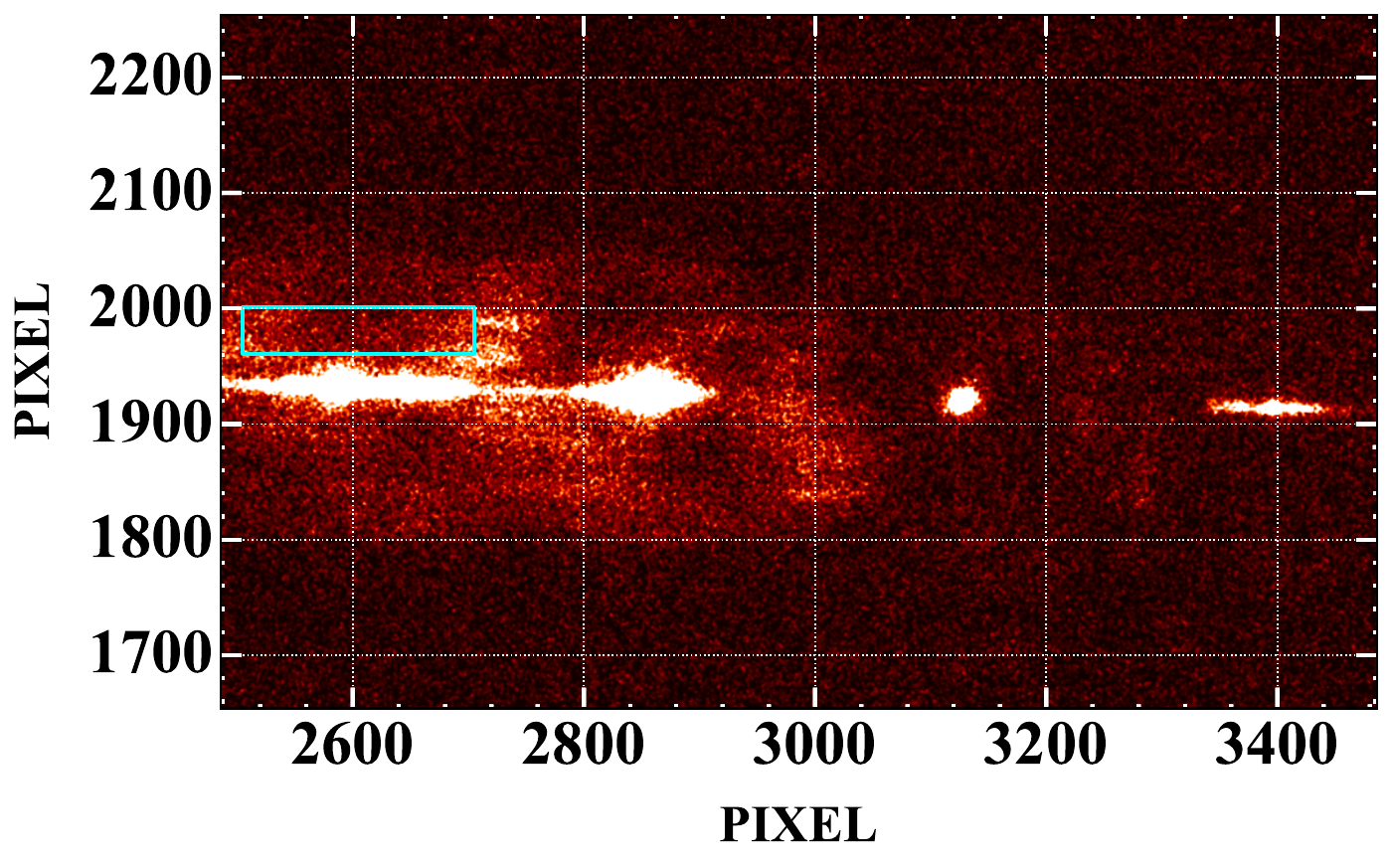}
\caption{FUV-G1 image of NGC~4151. The background can be seen to be significantly contaminated by the emission from the spiral arms. The cyan rectangular box shows the background extraction region for FUV-G1 in obs4.
\label{fig:fuvg}}
\end{figure}

We extracted the count spectra from the FUV-G1 and FUV-G2 merged images for obs4 and obs5, for which grating data are available. We used a cross-dispersion width of 40 pixels in the $-2$ order in the FUV gratings to extract the source spectra. The background region for FUV-G1 images of both the observations (obs4 and obs5) is significantly contaminated with emissions from the spiral arms and the diffuse emission from the galaxy.  In Fig.~\ref{fig:fuvg}, we show the FUV-G1 image of obs4, where the emission in the background region is significantly contaminated. We investigated the image and the background spectra of the upper and lower adjacent regions to the source.  The region above the source seemed more suitable for extracting the background spectrum (cyan rectangular box in Fig.~\ref{fig:fuvg}). The adjacent regions close to the source in the FUV-G2 image are relatively contamination-free. Therefore, we extracted the background spectrum from a source-free region away from the diffuse galaxy emission.

\begin{figure*}[ht!]
 \epsscale{0.55}
\plotone{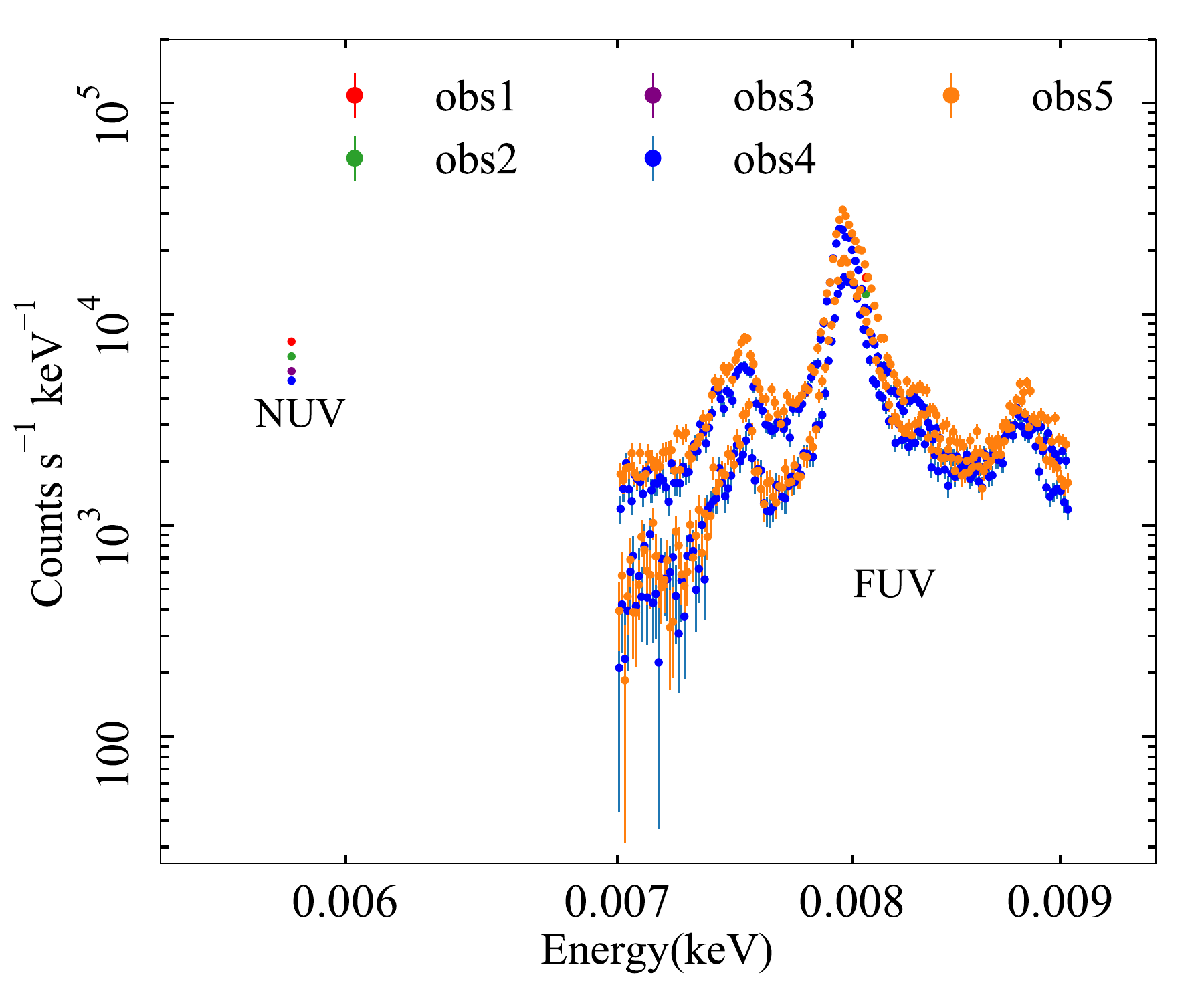}
 \epsscale{0.55}
\plotone{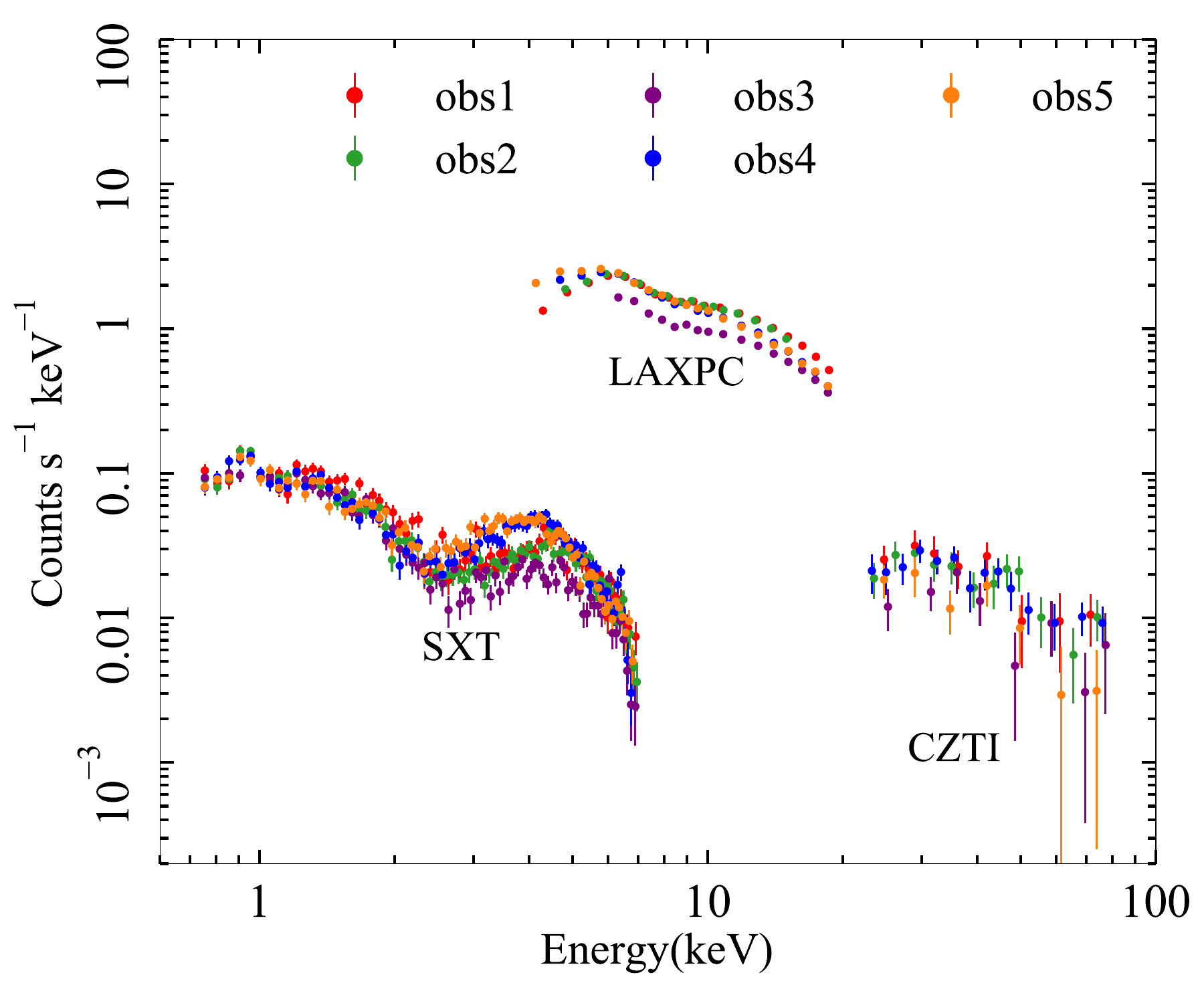}
\caption{The NUV photometric data and the FUV grating} (left), and SXT, LAXPC and CZTI (right) count spectra from all five observations.
\label{fig:countspec}
\end{figure*}

We show the count spectra from all five observations in Figure~\ref{fig:countspec}. The energy ranges covered by the data from different instruments are marked with shaded regions.
We performed the spectral analysis and model fitting with XSPEC (version 12.12.0; \citealt{arnaud1996xspec}) using $\chi ^2$ statistic for goodness of fit. We quote the errors at the 90\% confidence level unless otherwise specified.

\section{UVIT Spectral Analysis}
\label{sec:uvitspec}
We analyzed the UVIT spectral data (in the $5.5-9.5$ eV or $1305-2254$\angs{} band), namely the FUV-G1/FUV-G2 grating spectra,  and FUV-BaF$_2$,  NUV-B15 photometric flux from obs4 ($5.5-9.5$ eV or $1305-2254$\angs)  jointly and similarly for obs5 in the  $7-9.5$ eV ($1305-1770$\angs) band (as NUV data were not available). We used a variable cross-normalization constant between the gratings to account for any difference in flux measurements. 

We used the XSPEC model \redd~ \citep{cardelli1989relationship} to account for the reddening due to our Galaxy. We calculated the  color excess ($E(B-V)$)  using the following linear relation provided by \citet{guver2009relation}:

\begin{equation} \label{eq3:nhgal}
  N_{H}[{\rm~cm^{-2}}] = (2.21 \pm 0.09) \times 10^{21} A_{V}[{\rm mag}]
\end{equation}
where, $A_V = 3.1 \times E(B-V)$. We used $N_H = 2.07 \times 10^{20}~\rm cm^{-2}$ obtained from the  $N_H$ calculator available at the HEASARC website\footnote{\url{https://heasarc.gsfc.nasa.gov/cgi-bin/Tools/w3nh/w3nh.pl}}, and derived $E(B-V) = 0.03$.  We used this $E(B-V)$ and kept fixed in the \redd{}  model for the Galactic extinction in our spectral analysis.

We used a simple multi-temperature disk blackbody (\dbb) to model the underlying continuum and Gaussian lines for the BLR/NLR emission lines (e.g., \ion{Si}{4}/\ion{O}{4} $\lambda 1400$\angs, \ion{N}{4}] $\lambda 1486$\angs, \ion{C}{4}$\lambda 1549$\angs, \ion{He}{2} $\lambda 1640$\angs~ and \ion{O}{3}] $\lambda 1667$\angs). We observed some positive residuals near  $\sim1600$\angs~ and added a Gaussian line at the rest wavelength of  1599\angs. This improved the statistic by $\Delta \chi^2 = 8$ for two additional parameters: the line width and the normalization. This line could be the \ion{C}{4} satellite line observed during  $1980-1984$ with the International Ultraviolet Explorer ({\it IUE}) and was denoted by L2 \citep{ulrich1985narrow}. The narrow satellite emission lines of \ion{C}{4}, L1 (1515\angs), and L2 (1600\angs) are usually observed in the low flux states when the narrow emission lines become more prominent \citep{crenshaw2000space}. We used a Gaussian profile  (\texttt{GABS}) for the only absorption line near the rest wavelength of $1388$\angs~ with the best-fit values of width and strength being $1.4 \times 10^{-5}$ and $3.7 \times 10^{-5}$ keV, respectively. 

The semi-forbidden emission line \ion{C}{3}] and \ion{C}{2} lie around the edges of the NUV-B15 filter band. As the effective area around the edges decreases, we can treat the total NUV flux mostly due to the continuum. 

  \begin{figure*}
    \epsscale{0.55}
\plotone{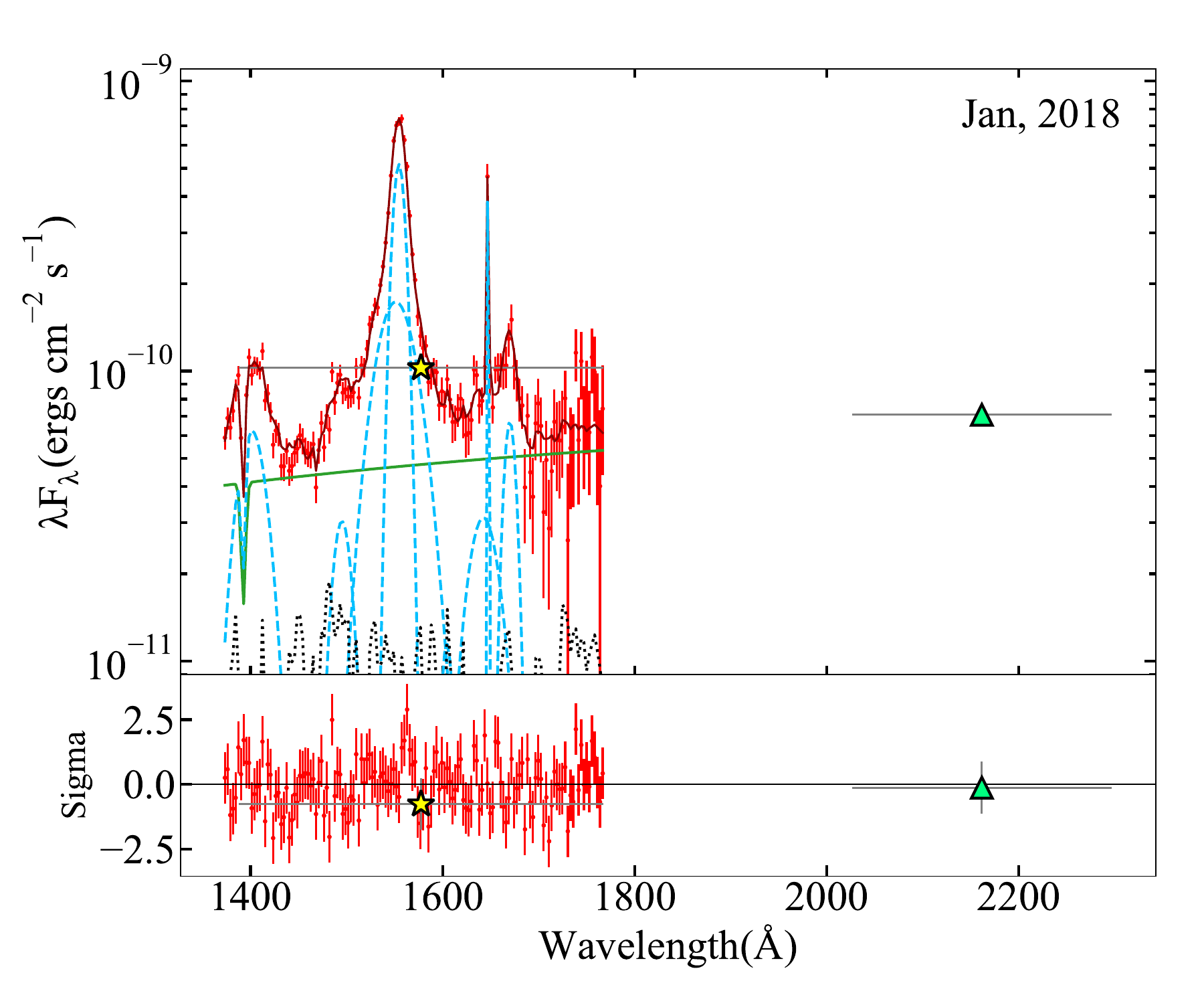}
    \epsscale{0.55}
\plotone{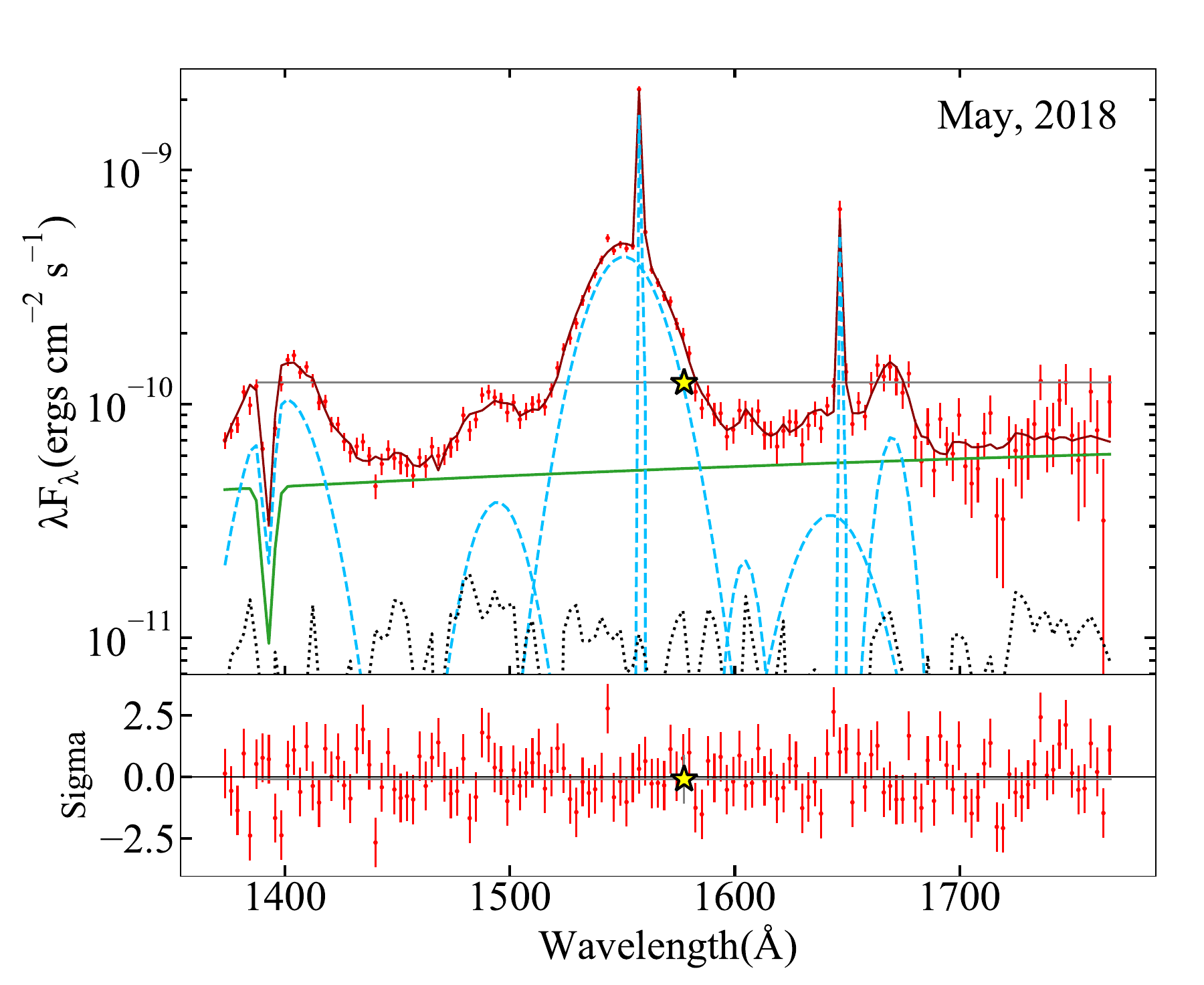}
\caption{Upper panels: For clarity, only the UVIT/FUV-G1  spectrum with the best-fit total model is shown (red). The model components are the intrinsic and Galactic absorbed power law (green solid line), Gaussian emission lines (cyan dashed line), and host galaxy template model (black dotted line).} Lower panels: (data-model)/error. The green triangle and yellow star show the NUV-B15 and FUV-BaF$_2$ photometric flux points.
\label{fig:uvit}
\end{figure*}

To account for the host galaxy contamination (spiral galaxy; \citealt{mahmoud2020discarding}), which is significant in our NUV band, we incorporated the Sb galaxy template  \citep{kinney1996template} in our fitting. Using optical observations with ground based telescopes, \citet{shapovalova2008long} calculated the host galaxy flux at 5100\AA~ as $\sim 1.1 \times 10^{-14}$\ergA{} for an aperture of $4.2 \times 19.8$ arcsec$^2$. We scaled the flux for our circular source extraction region of radius $8.3^{\prime\prime}$, assuming the host galaxy emission to be nearly uniform. We used the scaled galaxy template as a table model (hereafter sb\_temp) and fixed the normalization (total model flux in $5.5 - 10$ eV band ($\sim 1250-2250$\AA) is $6.5\times 10^{-12}$\ergS). For obs4, the contribution of the host galaxy emission to the total continuum emission is $0.8\%$ in the $5.5-10$ eV band.   

Previous observations have found this AGN, a Seyfert 1.5, to be highly obscured \citep{zdziarski1996origin,ricci2023changing}. Therefore, the UV emission may suffer high extinction. In the X-ray band, we observed a very high absorption column ($N_H \sim 10^{23} \rm cm^{-2}$) along the line of sight (discussed later); therefore, it is highly likely that the UV emission is also affected.  To account for the intrinsic extinction in the UV band, we created an XSPEC model (hereafter intr\_ext) using the empirical extinction curve provided by \citet{czerny2004extinction} for AGNs,

\begin{equation}
    \frac{A_\lambda}{E(B-V)} =  -1.36 + 13\log{\frac{1}{\lambda}}
    \label{eq1:intred}
\end{equation}
where, $\frac{1}{\lambda}$ ranges from 1.5 to 8.5 $\mu m^{-1}$.
We used the Balmer decrement (flux ratio of $H_\alpha$ and $H_\beta$) as a free parameter to estimate the intrinsic reddening (see Eq.~3 in  \citealt{2013ApJ...763..145D}). As NGC~4151 went through episodes of transition between high and low flux states, the flux ratio of the Balmer lines was also observed to vary. It has been observed that for the highest flux state, based on the continuum flux measured at 5100\angs, the Balmer decrement approaches $\sim 3.1$, while for the low flux state, this ratio is as high as $\sim 6$ \citep{shapovalova2010long,rakic2017intrinsic}. We compared the observed UVIT/FUV flux at 1440\angs~ with previous observations \citep{kraemer2006simultaneous,couto2016new} and found that it is similar to the low flux state observations. To determine the intrinsic extinction, we used a high spectral resolution ($ \sim 45800$) \textit{HST}/STIS-E140M (Space Telescope Imaging Spectrograph)  spectrum acquired during March 2000. The principal reason for choosing this data is that the observed flux at 1440\AA~ is similar to that in UVIT/FUV grating spectrum from obs4 taken in January 2018 ($\sim 3\times 10^{-14}$~\ergA). We compare the STIS and FUV spectra in Figure~\ref{fig:stis_spec} (left) of appendix \ref{sec:apendix}.

We used \kd, a relativistic accretion disc model available in XSPEC, as the continuum component and Gaussian profiles to account for the emission/absorption lines in the STIS spectrum. We obtained the best-fit value of the Balmer decrement, $4.95\pm 0.09$, for the intrinsic reddening. The details of the spectral analysis for STIS spectrum are provided in the appendix (\ref{sec:apendix}). We used this value of the Balmer decrement for the UVIT grating spectra and fixed it in our UV/X-ray joint fitting for each of the observations. 
 

The final model expression for the UVIT grating spectral analysis is \con~$\times$ \redd~$\times$ [sb\_temp + intr\_ext $\times$ (\dbb~+ emission lines) $\times$ absorption line]. In obs4, we obtained the best fit $\chi^2$ per degree of freedom (dof) = 308/274, and inner disk temperature, $kT_{in}=9.4_{-2.9}^{+4.7}$ eV. For obs5,  we obtained the best fit $\chi^2/$dof = 309/273, and  $kT_{in}=5.9_{-1.7}^{+4.0}$ eV. In Fig.~\ref{fig:uvit}, we show the data, best-fit model and model components, and the (data-model)/error for the UVIT grating spectra. We list the best-fit values of the emission line parameters in Table~\ref{tab:uvline}. The line centroids of all emission lines and the width of narrow or weak emission lines are fixed during error calculations.

\begin{deluxetable*}{clcccccccccc}
\tablenum{2}
\tablecaption{Best-fit parameters of the emission lines in the UV grating spectra of obs4 and obs5. $\lambda_o$ is the observed wavelength in the unit of \AA. $v_{fwhm}$ is the full-width half maxima (FWHM) of the emission line in the unit of \kms. $f_{line}$ is the line flux in the unit of $\rm photon~cm^{-2}~s^{-1}$. (f) -- fixed during the error calculation.} 
\label{tab:uvline}
\tablewidth{800pt}
\tablehead{
 & &\ion{Si}{4}/\ion{O}{4} & \ion{N}{4}] & \multicolumn3c {\ion{C}{4}} & \multicolumn2c{\ion{He}{2}} & \ion{O}{3}] \\
 \cmidrule(lr){5-7} \cmidrule(lr){8-9}
 &&&&Broad&Narrow&L2&Broad&Narrow&
} 
\startdata
\multirow{1}{*}{\begin{tabular}{c}obs4 
                  \end{tabular}}
& $\lambda_o$ (f)& $1396$ & $1490$ & $1546$ & $1549$ & $1599$ & $1637$ & $1642$ & $1665$ \\
&$v_{fwhm}$ & $7700_{-600}^{+640}$ & $4832_{-2046}^{+1266}$ & $9744_{-512}^{+613}$ & $3063_{-250}^{+242}$ & $2742$ (f) & $7086_{-2873}^{+1761}$ & $218$ (f) & $2735_{-909}^{+960}$ \\
&$f_{line}$ & $4.8_{-0.5}^{+0.5}$ & $1.4_{-0.2}^{+0.2}$ & $15_{-1}^{+1}$ & $14_{-1}^{+1}$ & $0.27_{-0.18}^{+0.17}$ & $1.8_{-0.4}^{+0.4}$ & $1.6_{-0.2}^{+0.2}$ & $1.5_{-0.2}^{+0.2}$ \\
\multirow{1}{*}{\begin{tabular}{c}obs5 
                  \end{tabular}}
& $\lambda_o$ (f)& $1394$ & $1490$ & $1546$ & $1553$ &$1599$  & $1637$ & $1642$ & $1665$ \\
&$v_{fwhm}$ & $10076_{-480}^{+515}$ & $8091_{-1178}^{+1594}$ & $10212_{-174}^{+180}$ & $289$ (f) & $2743$ (f) & $9727_{-1691}^{+2686}$ & $305$ (f) & $3770_{-888}^{+879}$ \\
&$f_{line}$ & $7.7_{-0.5}^{+0.5}$ & $2.1_{-0.2}^{+0.2}$ & $28_{-1}^{+1}$ & $8.1_{-0.8}^{+0.8}$ & $0.5_{-0.2}^{+0.2}$ & $1.95_{-0.39}^{+0.39}$ & $2.21_{-0.25}^{+0.25}$ & $1.6_{-0.2}^{+0.2}$ \\
\enddata
\end{deluxetable*}

\begin{deluxetable*}{clccccc}
\tablenum{3}
\tablecaption{Best-fit parameters of the model fitted only to the X-ray spectrum ($0.7-80$ keV; SXT/LAXPC/CZTI data). $N_H$ and $N_{H}^{WA}$ are in the unit of $\rm cm^{-2}$, $\xi$ is in $\rm erg~cm~s^{-1}$. The \dbb~ Norm is given by $(r_{in}/D_{10})^2 cos\theta$, where $r_{in}$ is the inner disk radius, $D_{10}$ is the distance in the unit of 10 kpc, and $\theta$ is the inclination angle. Emission lines are not shown here. }
\label{tab:sxt}
\tablewidth{700pt}
\tablehead{
\colhead{Model}&\colhead{Parameters} &\colhead{obs1} &   \colhead{obs2} & \colhead{obs3} &   \colhead{obs4} & \colhead{obs5}
}
\startdata
\multirow{1}{*}{\begin{tabular}{c}\xabs 
                  \end{tabular}}
&$\log \xi$& $<1.0$ & $-0.51$(f) & -- & -- & --\\
& $N_{H}^{WA}$($10^{22}{\rm~cm^{-2}})$ & $3.5_{-1.6}^{+3.2}$ & $10.4_{-4.8}^{+2.5}$& -- & -- & --\\
&$f_{c}^{XABS}$ & $0.67_{-0.02}^{+0.03}$ & $0.76_{-0.15}^{+0.07}$& -- & -- & --\\
\multirow{1}{*}{\begin{tabular}{c}\tbp 
                  \end{tabular}}
& $N_H$($10^{22}{\rm~cm^{-2}})$& $32.5_{-2.3}^{+2.2}$ & $31.1_{-3.0}^{+1.9}$ & $26.7_{-2.2}^{+2.2}$ & $17.1_{-1.0}^{+1.1}$ & $12.2_{-0.8}^{+0.8}$ \\
& $f_{c}^{TBPCF}$ & $0.92_{-0.02}^{+0.01}$ & $0.92_{-0.03}^{+0.03}$ & $0.93_{-0.01}^{+0.01}$ & $0.957_{-0.004}^{+0.004}$ & $0.941_{-0.006}^{+0.005}$\\
\zbb & $kT$ (keV) & 0.09 (f) & 0.09 (f) & 0.09 (f) & 0.09 (f)& 0.09 (f)\\
&Norm ($ 10^{-5}$) &$<10.1$& $<7.3$& $7.8_{-3.2}^{+3.2}$& $<6.3$ & $<8.5$  \\
\iref & $f_{refl}$& 0.3(f) & 0.3(f)& 0.3(f) & 0.3(f)& 0.3(f)\\
& $\xi$ & 0(f) &  0(f)& 0(f) &0(f)& 0(f)\\
\multirow{1}{*}{\begin{tabular}{c}\thc
                  \end{tabular}}
& $\Gamma$ & $1.70_{-0.05}^{+0.05}$ & $1.85_{-0.07}^{+0.08}$ & $1.59_{-0.04}^{+0.04}$& $1.70_{-0.03}^{+0.03}$ & $1.65_{-0.03}^{+0.03}$\\ 
&$f_c^{THCOMP}$& 1(f)&1(f)&1(f)&1(f)&1(f)\\
& $kT_e$ (keV)  & 100(f) & 100(f)&100(f) & 100(f) & 100(f)\\
\dbb & $kT_{in}$ (eV) & 5 (f) &  5 (f)& 5 (f)& 5 (f)& 5 (f)\\
&Norm ($10^{9}$) & $4.0_{-1.2}^{+1.8}$ & $10.6_{-4.4}^{+10.1}$ & $0.7_{-0.2}^{+0.2}$ & $2.6_{-0.5}^{+0.7}$ & $1.5_{-0.3}^{+0.4}$\\
\enddata
\end{deluxetable*}

\section{X-ray Spectral Analysis}
\label{sec:xrayspec}

We analyzed the SXT, LAXPC, and CZTI spectral data jointly for each observation. We used the  $0.7-7$ keV  SXT data, $3-20$ keV  LAXPC data, and $22-80$ keV  CZTI data. We began the analysis with only the SXT data with a simple power law model modified with the Galactic absorption. For the Galactic X-ray absorption, we used \tba~ and fixed the  equivalent hydrogen column density $N_H$ at $2.07\times 10^{20}\rm cm^{-2}$ obtained from the HEASoft $N_H$ calculator\footnote{\url{https://heasarc.gsfc.nasa.gov/cgi-bin/Tools/w3nh/w3nh.pl}}.  We also used a neutral partial covering absorption model available in XSPEC \tbp~ to account for the absorbed primary continuum and unabsorbed scattered continuum. For all five observations, we detected the presence of narrow emission lines due to Fe K$\alpha$ at 6.4 keV, He-like \ion{Ne}{9} at 0.92 keV. We also observed positive residuals around 1.8 keV. We added a narrow ($\sigma =1$ eV) Gaussian emission line fixed at 1.84 keV. This improved the fit by  $\Delta \chi^2=5$ (obs1), 6 (obs2), 3 (obs3 and obs4), and 8 (obs5) for one additional free parameter, the normalization. This line is most likely the blend of Si K and triplet lines at $\sim 1.84\kev$ due to He-like Si i.e., \ion{Si}{13} that have been detected in the \chandra{}/HETG spectrum by  
\citep{kraemer2020mass}. Therefore,  we kept the Gaussian line at $1.84\kev$ included in our spectral fitting. Removing the line does not change the derived continuum parameters in any significant way.

In obs3, we found excess emission at around six keV.  Adding a narrow ($\sigma=1$~eV) Gaussian line at 6.01~keV improved the $\chi^2$ by 28 for two additional parameters, namely, the line centroid and normalization. 

We used \zbb~ to account for the soft excess emission. For obs3, this improved the $\chi^2$ by 15 for two additional parameters with the best-fit value of $kT \sim 0.09$~keV.  We fixed the $kT$ at 0.09 keV, as the temperature is similar in all the observations, and let the normalization vary for the other observations. For these, including a soft excess component resulted in no significant improvement in the fit.  We could constrain the normalization in obs3 and obtain an upper limit for the remaining four observations. The upper limits for the normalization are consistent with that of obs3. 

The spectral model used for the SXT data can be expressed as \tba $\times$ [\zbb + \tbp $\times$ (\zpo + Fe~K$\alpha$ + Si~K$\alpha$  + \ion{Ne}{9})]. We obtained the best fit $\chi^2/dof$ of 108/72 (obs1), 85/72 (obs2), 68/68 (obs3), 79/74 (obs4), and 80/73 (obs5).  We obtained a range for the  unabsorbed $2-10$ keV primary X-ray continuum flux of $1.1-2.1\times10^{-10}~\rm erg~cm^{-2}~s^{-1}$. This is similar to that observed when the source is in a bright X-ray state \citep{2010MNRAS.408.1851L}.

Next, we included the LAXPC and CZTI spectral data. We used a model systematic error of $3\%$ to account for calibration issues. We replaced the simple power-law component with a thermal comptonization model \thc~ \citep{zdziarski2020spectral}, which we convolved with the \dbb~ model. The parameters for the \thc~ are the photon index ($\Gamma$), high energy electron temperature ($kT_e$), the fraction of seed photons being Comptonized ($f_{c}^{THCOMP}$), and the redshift. This model can Comptonize the disk seed photons to generate the soft excess emission or the hard X-ray continuum depending upon the electron temperature of the Comptonizing medium. Here, we fixed the electron temperature at 100 keV, usually observed in the bright state \citep{2010MNRAS.408.1851L}. We fixed the $f_{c}^{THCOMP}$ at 1. Additionally, we used a cross-normalization constant for the LAXPC and CZTI spectral data. For the CZTI spectra, we obtained similar values for this constant in all the observations. Therefore, we fixed this to the best-fit value of $0.77$, as obtained in obs2 and obs4, in all the observations. For the LAXPC, we found this constant to be $0.86$ (obs1, obs2, and obs4) and $1.1$ (obs3 and obs5), which we fixed during the subsequent analyses. We also fixed the inner disk temperature ($kT_{in}$) in \dbb~ at 5 eV and varied the normalization only since changing the $kT_{in}$ from 5 to 50 eV did not affect the X-ray continuum parameters. 

The ubiquitous presence of a narrow Fe~K$\alpha$ emission line in all the observations indicates the signature of distant reflection. We used a convolution model \iref~ \citep{magdziarz1995angle} to model this reflection component. We extended the energy grid from $10^{-4}$ to 1000 keV to use the convolution models. 
The parameters of this model are $f_{refl}$ (reflection fraction), the metal abundance, Fe abundance, inclination angle, disk temperature, and $\xi$ (disk ionization parameter). Initially, we varied the reflection fraction for all the observations, but we could constrain this parameter for obs4 and obs5 only. For obs4, it lies between $0.03-0.4$, and for obs5, it lies between $0.3-0.9$. We fixed this parameter at 0.3 for all the observations, assuming the distant reflection to be nearly constant over time. The metal and iron abundances and inclination angle are fixed at solar and $45^\circ$, respectively. We fixed the ionization parameter to 0, assuming the distant reflector to be neutral. 
Adding the reflection component improves the fit by $\Delta \chi^2 \sim 5-20$ for no additional free parameter in four observations (except for obs1).  We obtained the reduced $\chi^2$ of 1.8 (obs1), 1.2 (obs2), 1.1 (obs3), 0.9 (obs4), and 0.9 (obs5). 

\begin{figure}[ht!]
 \epsscale{1.25}
\plotone{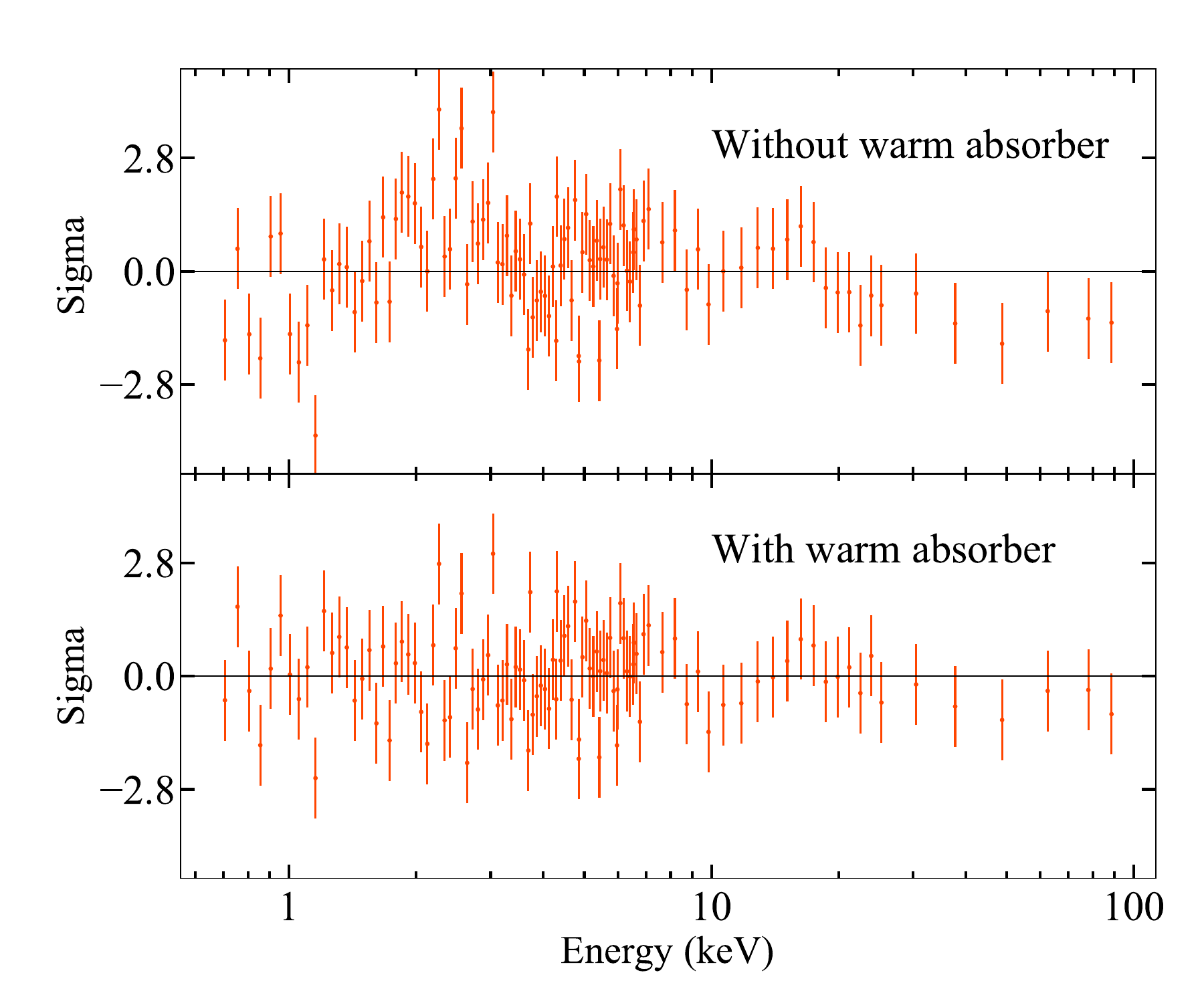}
\caption{Obs1: The (data-model)/error before (top) and after (bottom) using the XABS model component.  
\label{fig:xabs_obs1}}
\end{figure}

We observed significant residuals in obs1 around 1~keV. We tested the presence of warm absorbers in this observation using the SPEX \citep{kaastra1996spex} model \xabs~ \citep{steenbrugge2003xmm} available as a table model for XSPEC \citep{parker2019x}. The parameters of this model are the logarithm of the ionization parameter ($\log \xi$), hydrogen column density ($N_{H}^{WA}$), root mean square (rms) velocity of the absorbing plasma (v), covering factor of the absorbing plasma ($f_{c}^{XABS}$) and the redshift. We used this table model for the X-ray spectra. This improved the $\chi^2$ by 70 for four additional parameters. The data could not constrain the rms velocity; therefore, we fixed it to 200 \kms, which did not change the $\chi^2$. The best-fit values of \xabs~ parameters are: $\log \xi < 1.02$,  $N_{H}^{WA} \sim 3.6 \times 10^{22}~\rm cm^{-2}$, $f_{c}^{XABS} \sim 0.7$. The fit residuals before and after adding the \xabs{} component are shown in Figure~\ref{fig:xabs_obs1}.  

We also tested the presence of warm absorbers in all other observations. The $\log \xi$ is fixed to the best-fit value obtained from obs1 ($-0.51$). We fixed the turbulent velocity arbitrarily at  200 \kms. For obs2, adding this component improved the $\chi^2$ by 33 for three additional parameters.  The best-fit parameters are listed in Table~\ref{tab:sxt}. Adding this model component did not improve the fit statistics significantly for obs3, obs4, and obs5. 
We tested the variability of the column density of the warm absorber by fixing the covering fraction to 0.7. 
We could only obtain the $90\%$ upper limit of $5\times10^{21}{\rm~cm^{-2}}$ for obs3,  $4\times 10^{21}{\rm~cm^{-2}}$ for obs4, and $2\times 10^{21}{\rm~cm^{-2}}$ for obs5.   This could be due to the variation in the warm absorber along the line of sight, as observed earlier in this source \citep{schurch2002characterizing,zoghbi2019revisiting}. 
Therefore, we did not use this component for these observations. The final reduced $\chi^2$ for obs1 and obs2 are 1.2 and 0.9, respectively. The final expression of the combination of model is,  \con~ $\times$ \tba~ $\times$ [\zbb~ + \xabs~ $\times$ \tbp $\times$ (\iref$\times$\thc$\times$\dbb~ + Fe~K$\alpha$ + Si~K$\alpha$ + \ion{Ne}{9})].
%
The $N_H$ in the \tbp~ component varies from $\sim 1.2-3.4 \times 10^{23}~\rm cm^{-2}$ and the $\Gamma$ varies from $\sim 1.56-1.87$.  Next, we proceeded to the UV/X-ray broadband spectral modeling.

\begin{figure*}
\centering
\gridline{\fig{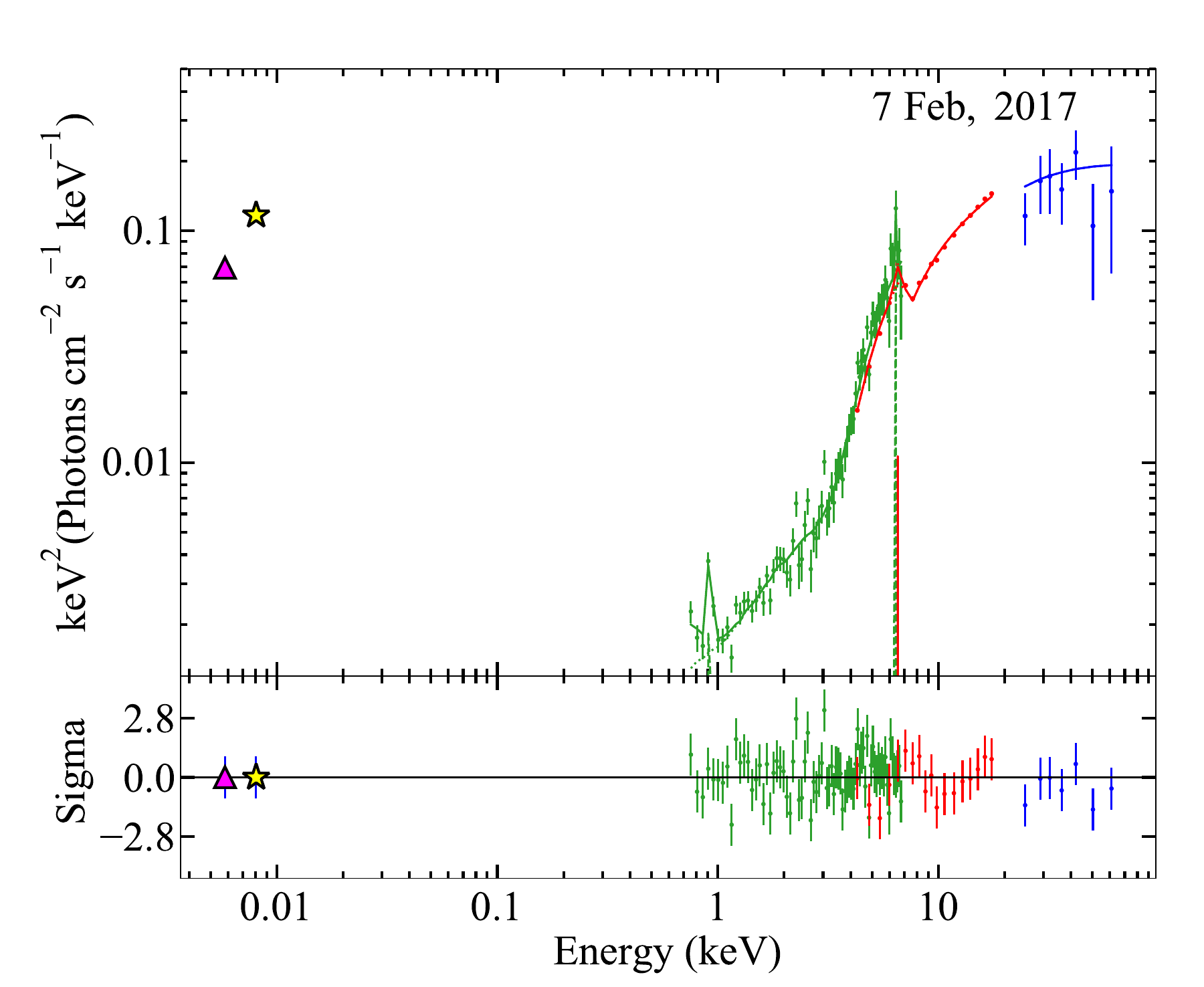}{0.45\textwidth}{(a)}
           \fig{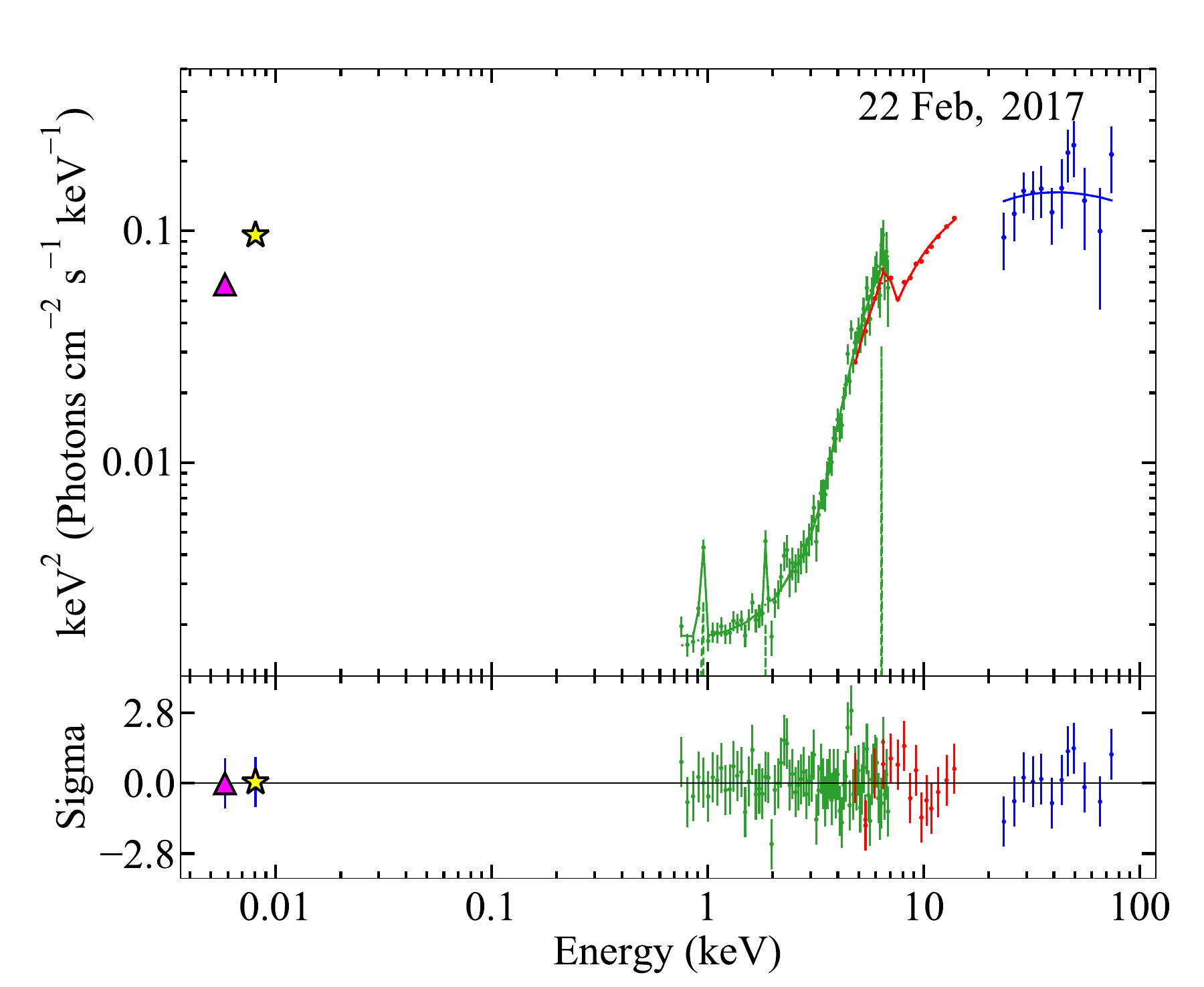}{0.45\textwidth}{(b)}
          }
\gridline{\fig{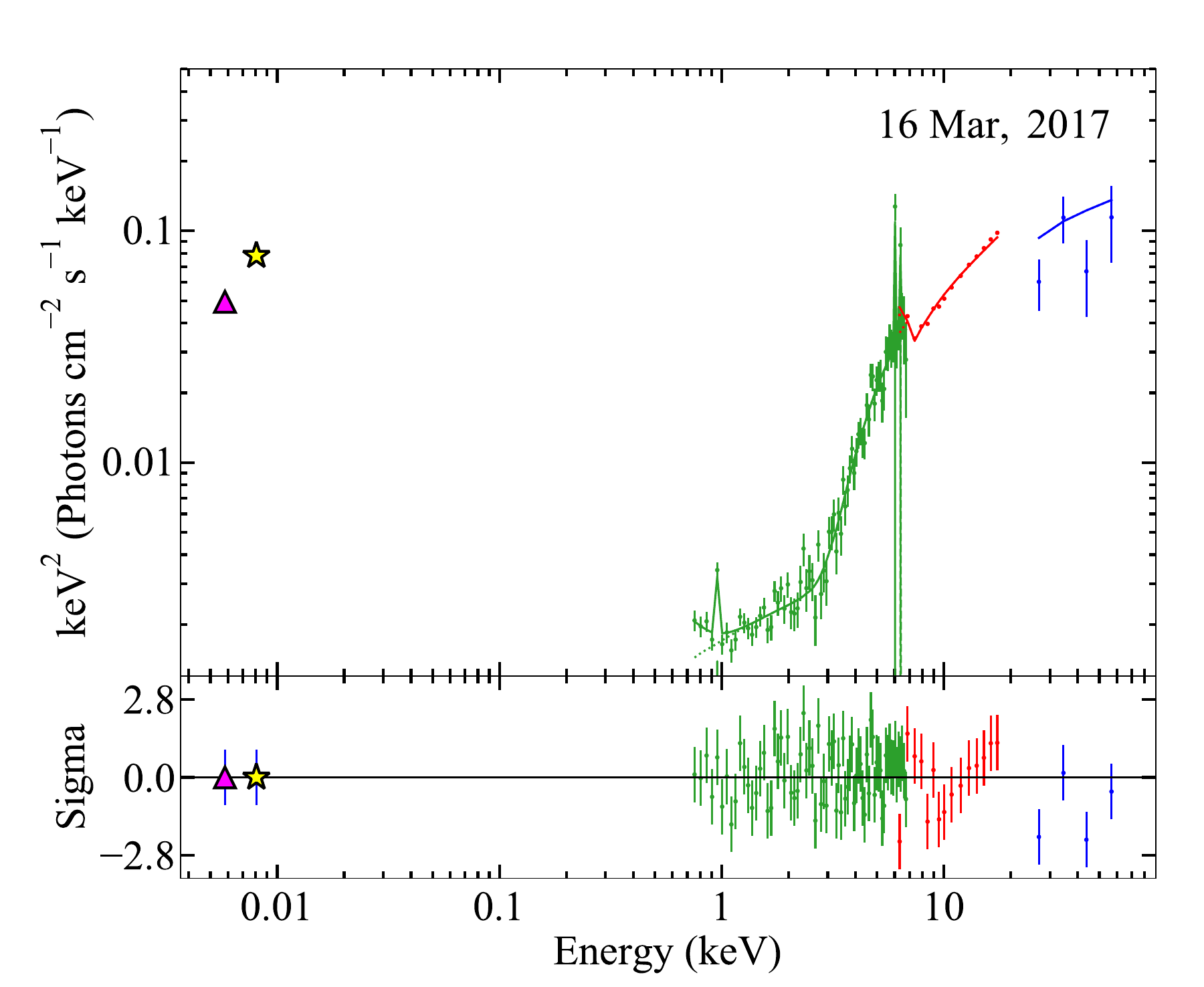}{0.45\textwidth}{(c)}
           \fig{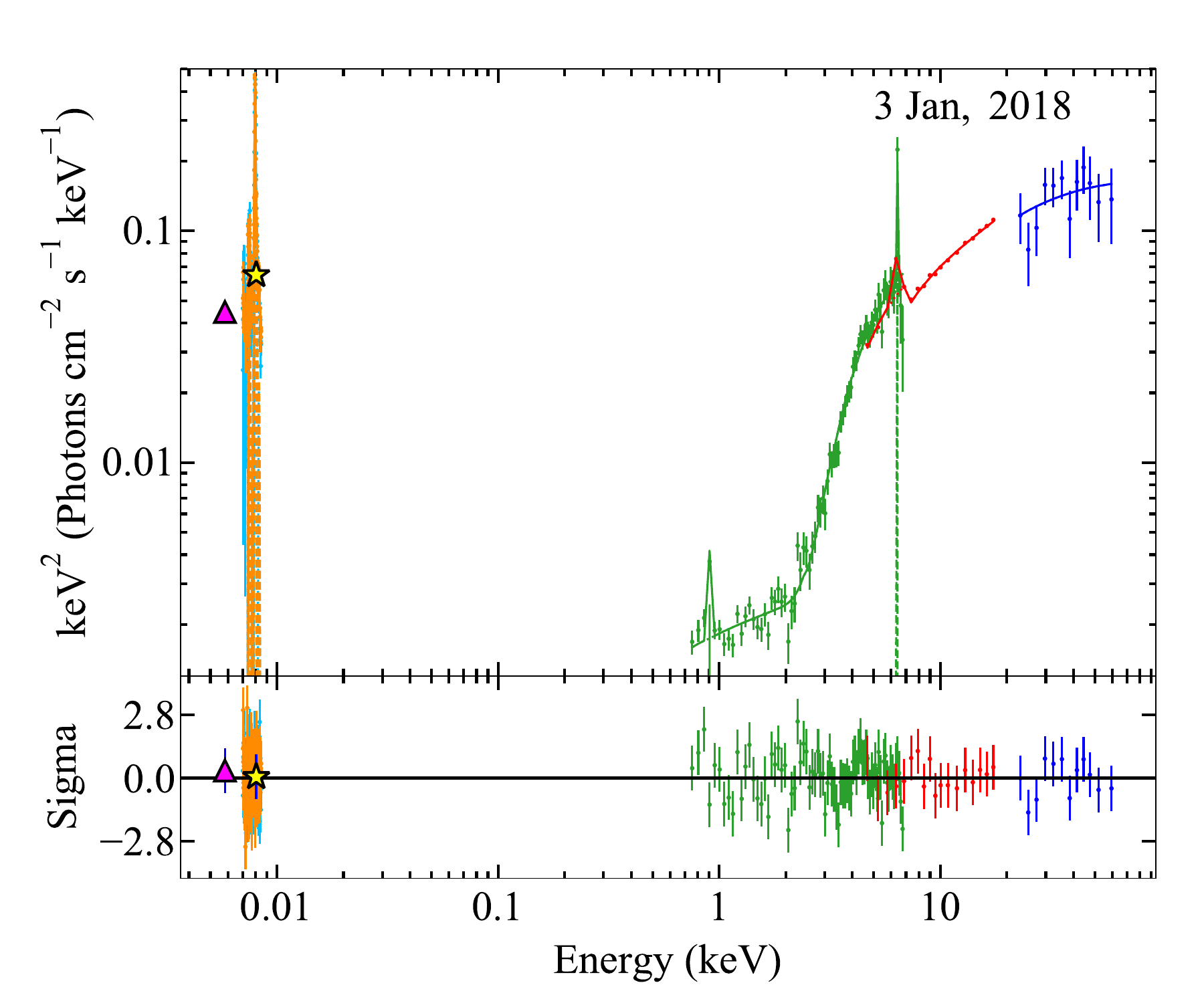}{0.45\textwidth}{(d)}
          }
\gridline{\fig{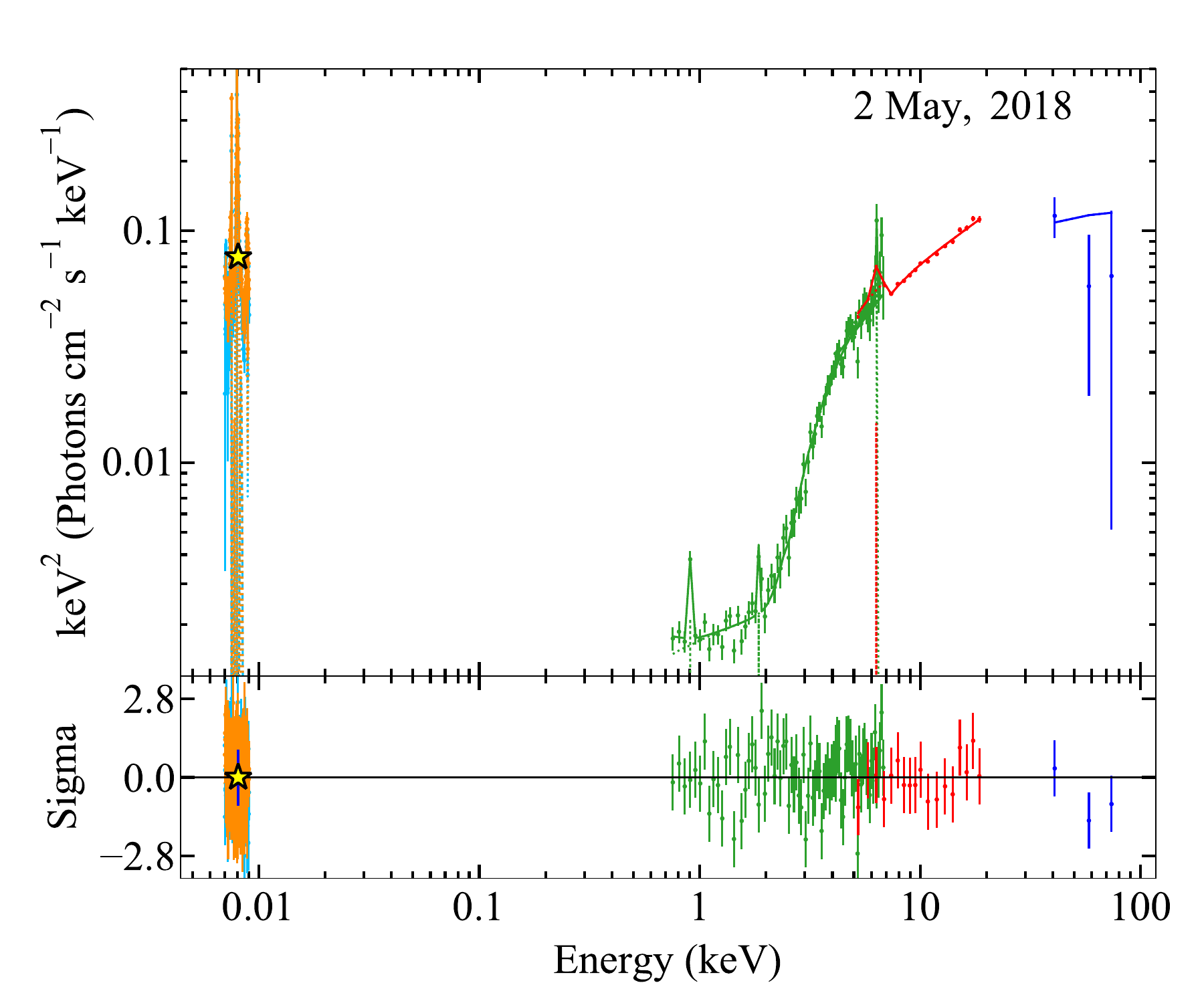}{0.45\textwidth}{(e)}
          }
\caption{UV and X-ray joint final best-fit model for the five observations. For each figure, the upper panel shows the best-fit model and data. The dashed lines show the emission line components.  The lower panel shows the  (data-model)/error.  The data are rebinned for plotting purposes. UVIT/FUV-G1 and FUV-G2 spectra are in orange and cyan, NUV-B15 and FUV-BaF$_2$ data points are represented by the triangle and star symbols, respectively, SXT spectrum in green, LAXPC spectrum in red, and  CZTI spectrum in blue.}
\label{fig:joint_o}
\end{figure*}

  \begin{figure}[ht!]
 \epsscale{1.25}
\plotone{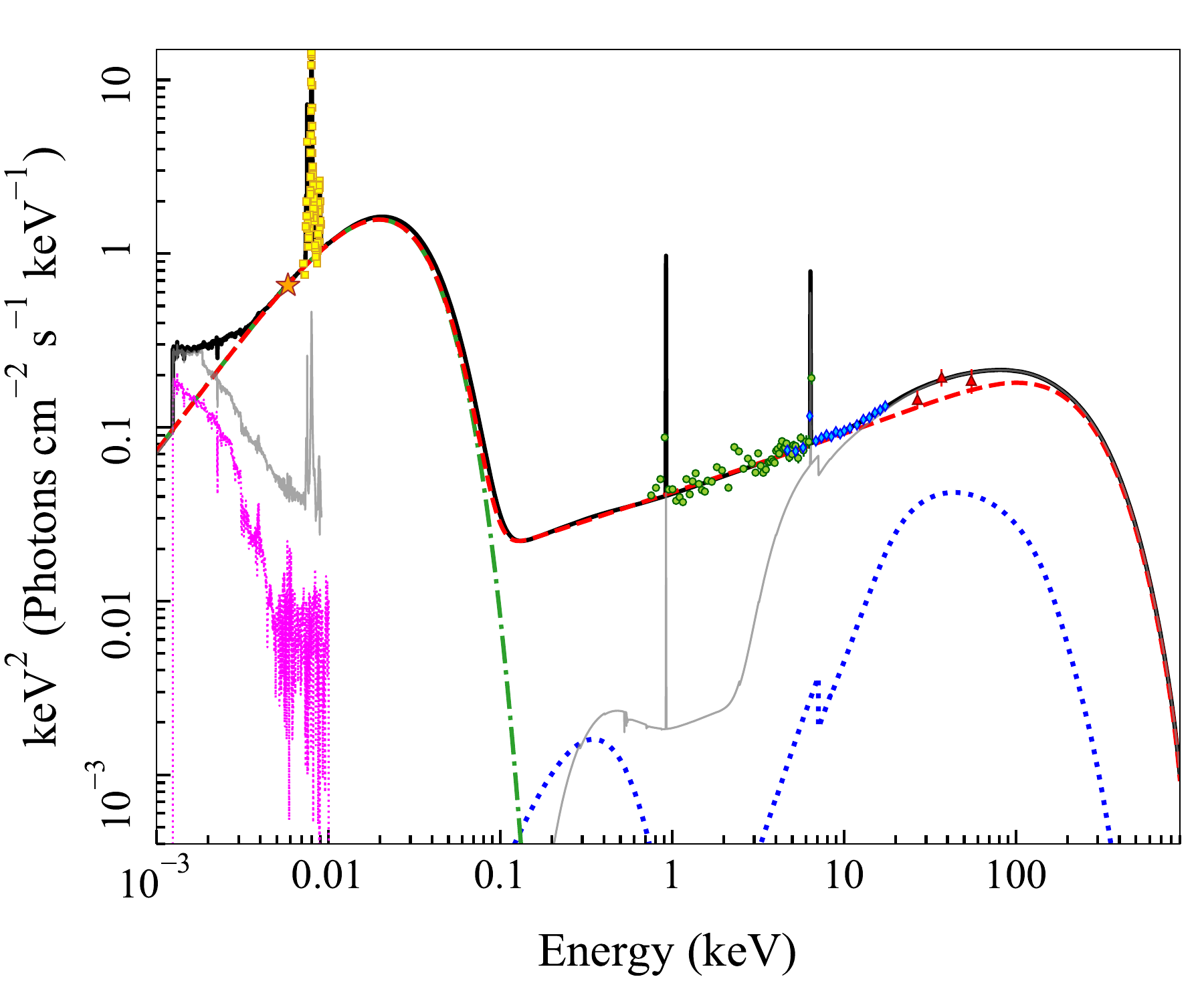}
\caption{The black solid line shows the final best fit unabsorbed total model and the gray is the same for the absorbed model. The different unabsorbed components are shown by --  green dash-dotted: accretion disk emission; red dashed: comptonized disk emission; blue dotted: soft excess and the Compton hump (reflection emission); fuchsia: host galaxy (sb\_temp) template. The dereddened and unabsorbed spectral data are shown by the orange star (NUV-B15), yellow squares (FUV-G1), green circles (SXT), blue diamonds (LAXPC), and red triangles (CZTI).  
\label{fig:abs_unabs_c}}
\end{figure}

\section{UV/X-ray Joint Spectral Analysis}
\label{sec:uvXrayspec}

We started our UV/X-ray joint spectral analysis with obs4 and obs5, for which the FUV-G1 and FUV-G2 grating spectral data are available. We also used the FUV-BaF$_2$ and NUV-B15 single-channel spectral data wherever available. We put together the final models fitted separately to the UV and X-ray spectral datasets, our joint UV -- X-ray final model expression reads as \con~ $\times$ \tba~ $\times$ \redd~ $\times$ \big[sb\_temp + intr\_ext $\times$ absorption line $\times$ \{\zbb~ + \xabs~ $\times$ \tbp $\times$ (\iref$\times$\thc$\times$\dbb~ + Fe~K$\alpha$ + Si~K$\alpha$ + \ion{Ne}{9} + UV Emission lines)\}\big]. The \xabs~ model is used only for the obs1 and obs2. 

As before, we used relative normalizations for the spectral data from different instruments. We fixed the relative normalizations at 1 for the UVIT/FUV-G1, FUV-BaF$_2$, NUV-B15, and SXT data. For the rest of the instruments, we fixed the relative normalizations at the values obtained earlier from the UVIT, and X-ray spectral analysis performed separately.
We fixed the E(B-V) in \redd~ and Balmer decrement in intr\_red  at 0 and 2.72, respectively, for the X-ray spectral data. We also fixed the partial covering fraction ($f_{c}^{TBPCF}$) and $N_H$ in \tbp, $N_H$ in \tba, $N_{H}^{WA}$ and $f_{c}^{XABS}$ in \xabs{}  at 0 for the UV spectral data.

Initially, we varied the \dbb{} temperature and normalizations. We found that the best-fit normalizations for obs4 and obs5 were similar within errors, while we could not constrain the normalizations in the case of obs1, obs2, and obs3 as only two UV data points from the broadband filters are available for these observations. Therefore, we fixed the \dbb{} normalizations for obs1, obs2, obs3, and obs5 to the best-fit value derived for obs4.   
We varied the $\Gamma$ and the $f_{c}^{THCOMP}$ in \thc; the $N_H$ and partial covering in \tbp; $\log \xi$, $N_{H}^{WA}$ and $f_{c}^{XABS}$ in \xabs; and the normalization of the \zbb~ and emission lines in X-ray bands.
For obs2, we fixed the $\log \xi$ at the best-fit value of $-0.51$ as obtained in obs1. 
 We varied the normalizations of the UV emission lines and found them similar to those obtained during the separate UVIT spectral analysis. Therefore, we fixed all the UV emission or absorption line parameters.
 
 We obtained  a $\chi^2/dof = 401.2/395$ for  obs4 and $400.4/395$ for  obs5. The total unabsorbed flux from the UV emission lines has increased by a factor of $\sim 1.3$ in obs5  compared to those in obs4.
 
For obs1, obs2, and obs3, we have no information regarding the emission lines in FUV/NUV bands. 
We arbitrarily used the emission lines obtained from obs5 as a reference for the FUV band in obs1, obs2, and obs3. For obs1 and obs2, we used a constant multiplicative factor for the FUV emission lines to account for the increase in the line flux (if any) with the rise of the total UV flux. we obtained the best-fit value for this constant $\sim 1.23$ in obs2 and $\sim 1.47$ in obs1, assuming the same intrinsic reddening suffered during all the observations. 
The best-fit values are listed in Table~\ref{tab:uvsxtsb}. The data, the best-fit model, and the residuals in terms of (data-model)/error for all observations are shown in Fig.~\ref{fig:joint_o}. For obs4, the final unabsorbed and de-reddened SED and its different model components are shown in Figure~\ref{fig:abs_unabs_c}. 

\begin{deluxetable*}{clccccc}
\tablenum{4}
\tablecaption{Best-fit parameters of the final model fitted to the joint UV -- X-ray spectra (UVIT/SXT/LAXPC/CZTI). The normalizations of the X-ray emission lines are in the unit of $\rm photons~cm^{-2}~s^{-1}$.  }
\label{tab:uvsxtsb}
\tablewidth{700pt}
\tablehead{
\colhead{Model}&\colhead{Parameters} &\colhead{obs1} &   \colhead{obs2} & \colhead{obs3} &   \colhead{obs4} & \colhead{obs5}
}
\startdata
\multirow{1}{*}{\begin{tabular}{c}\xabs 
                  \end{tabular}}
&$\log \xi$& $<0.7$ & $-0.51$(f)& -- & -- & --\\
& $N_{H}^{WA}~\times 10^{22}$ & $2.8_{-1.6}^{+1.9}$ & $10.4_{-4.2}^{+2.4}$& -- & -- & --\\
&$f_{c}^{XABS}$ & $0.72_{-0.04}^{+0.07}$ & $0.8_{-0.1}^{+0.1}$& -- & -- & --\\
\multirow{1}{*}{\begin{tabular}{c}\tbp 
                  \end{tabular}}
& $N_H~\times 10^{22}$& $33.9_{-2.0}^{+3.8}$ & $32.3_{-2.8}^{+3.3}$ & $22.8_{-1.7}^{+1.9}$ & $16.6_{-0.9}^{+0.9}$ & $12.1_{-0.7}^{+0.8}$ \\
& $f_{c}^{TBPCF}$ & $0.926_{-0.014}^{+0.013}$ & $0.92_{-0.03}^{+0.03}$ & $0.914_{-0.011}^{+0.010}$ & $0.954_{-0.004}^{+0.004}$ & $0.944_{-0.005}^{+0.005}$\\
\iref & $f_{refl}$& $0.3$(f) & $0.3$(f) & $0.3$(f) & $0.3$(f)& $0.3$(f) \\
& $\xi$ & $0$(f) &  $0$(f)& $0$(f) &$0$(f)& 0(f)\\
\multirow{1}{*}{\begin{tabular}{c}\thc
                  \end{tabular}}
& $\Gamma$ & $1.74_{-0.06}^{+0.07}$ & $1.87_{-0.07}^{+0.08}$ & $1.56_{-0.05}^{+0.05}$& $1.67_{-0.03}^{+0.04}$ & $1.68_{-0.04}^{+0.04}$\\ 
&$f_{c}^{THCOMP}$& $0.008_{-0.002}^{+0.005}$ & $0.02_{-0.01}^{+0.01}$&$0.002_{-0.001}^{+0.001}$ & $0.007_{-0.001}^{+0.002}$& $0.005_{-0.001}^{+0.001}$\\
& $kT_e$ (keV) & 100(f) & 100(f)&100(f) & 100(f) & 100(f)\\
\dbb & $kT_{in}$ (eV)&  $10.3_{-0.2}^{+0.2}$ & $9.6_{-0.2}^{+0.2}$ & $9.0_{-0.2}^{+0.2}$& $8.7_{-2.0}^{+4.7}$ & $9.0_{-0.1}^{+0.1}$\\
& Norm ($10^{10}$) & $4.3$(f) & $4.3$(f) & $4.3$(f) & $4.3_{-2.1}^{+5.5}$ & $4.3$(f)\\
\zbb & $kT$ (keV) & 0.09 (f) & 0.09 (f) & 0.09 (f) & 0.09 (f)& 0.09 (f)\\
&Norm ($ 10^{-5}$) &$11.6_{-8.0}^{+7.7}$& $<4.5$& $8.0_{-4.2}^{+4.2}$& $<8.7$ & $<8.1$  \\
Fe K$\alpha$ &norm ($10^{-4}$)&  $2.9_{-1.5}^{+1.7}$ & $<3.4$ & $1.9_{-1.0}^{+1.1}$ & $5.0_{-1.2}^{+1.2}$ & $2.2_{-0.9}^{+0.9}$\\
\ion{Ne}{9}& norm ($10^{-3}$)& $5.4_{-2.6}^{+4.2}$ & $10.2_{-3.3}^{+4.2}$ & $0.9_{-0.5}^{+0.5}$ & $3.1_{-1.1}^{+1.2}$ & $2.5_{-0.9}^{+1.0}$ \\
Si K$\alpha$& norm ($10^{-4}$)& $<8.0$ & $8.8_{-7.1}^{+7.1}$ & $<4.1$ & $4.2_{-3.9}^{+4.2}$ & $6.9_{-3.8}^{+4.2}$  \\
$\chi^2/dof$ && 125.0/104  &89.1/105 & 109.3/104 & 401.2/395 & 400.4/395 
\enddata
\end{deluxetable*}

\section{Results and discussion}
\label{sec:res_disc}

We analyzed the  \asat~ multi-wavelength data from the five observations performed during 2017 Feb/March and 2018 Jan/May. In the UV band, we derived the intrinsic continuum after removing the effect of intrinsic and Galactic extinction and correcting for emission from the BLR/NLR. We obtained the inner disk temperature varying between $\sim 8.7-10.3$ eV. The variation in the disk temperature may not be real but due to our assumption of constant intrinsic reddening (see below). The emission line widths are of the order of a few hundred to a few thousand \kms, typical of the NLR and BLR emission. During our observations (obs4), the observed UV flux is near the minimum level (at 1350\angs, flux $\sim 4 \times 10^{-14}$\ergA) observed in the past as reported in \citet{kraemer2006simultaneous}. 

\begin{figure}[ht!]
 \epsscale{1.1}
\plotone{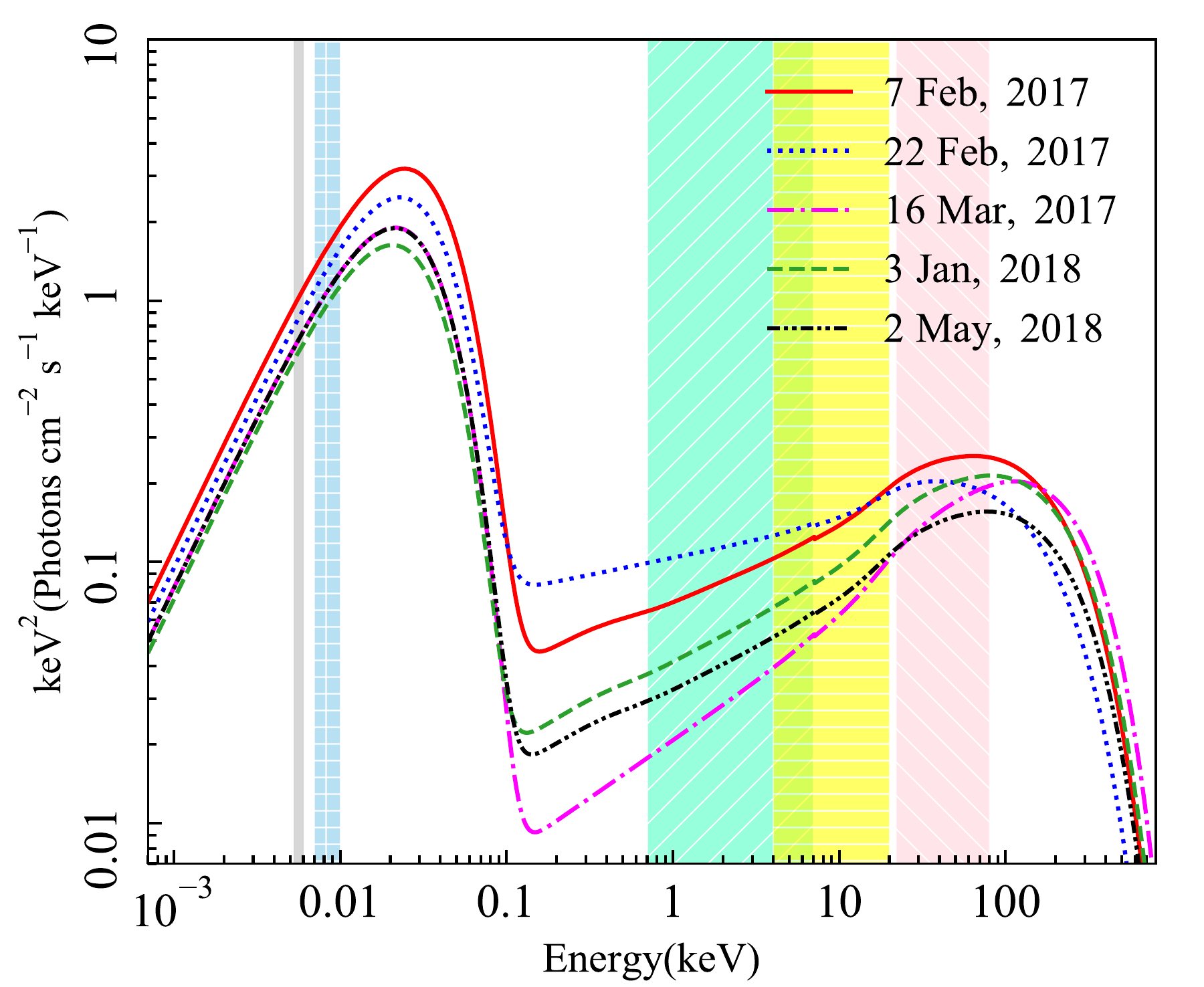}
\caption{Total unabsorbed continuum model for all the observations. The shaded energy bands show the NUV-B15 (gray), FUV-G1/G2 (sky blue), SXT (green), LAXPC (yellow), and CZTI (pink).   
\label{fig:all_eem}}
\end{figure}

 \begin{figure}[ht!]
 \epsscale{1.25}
\plotone{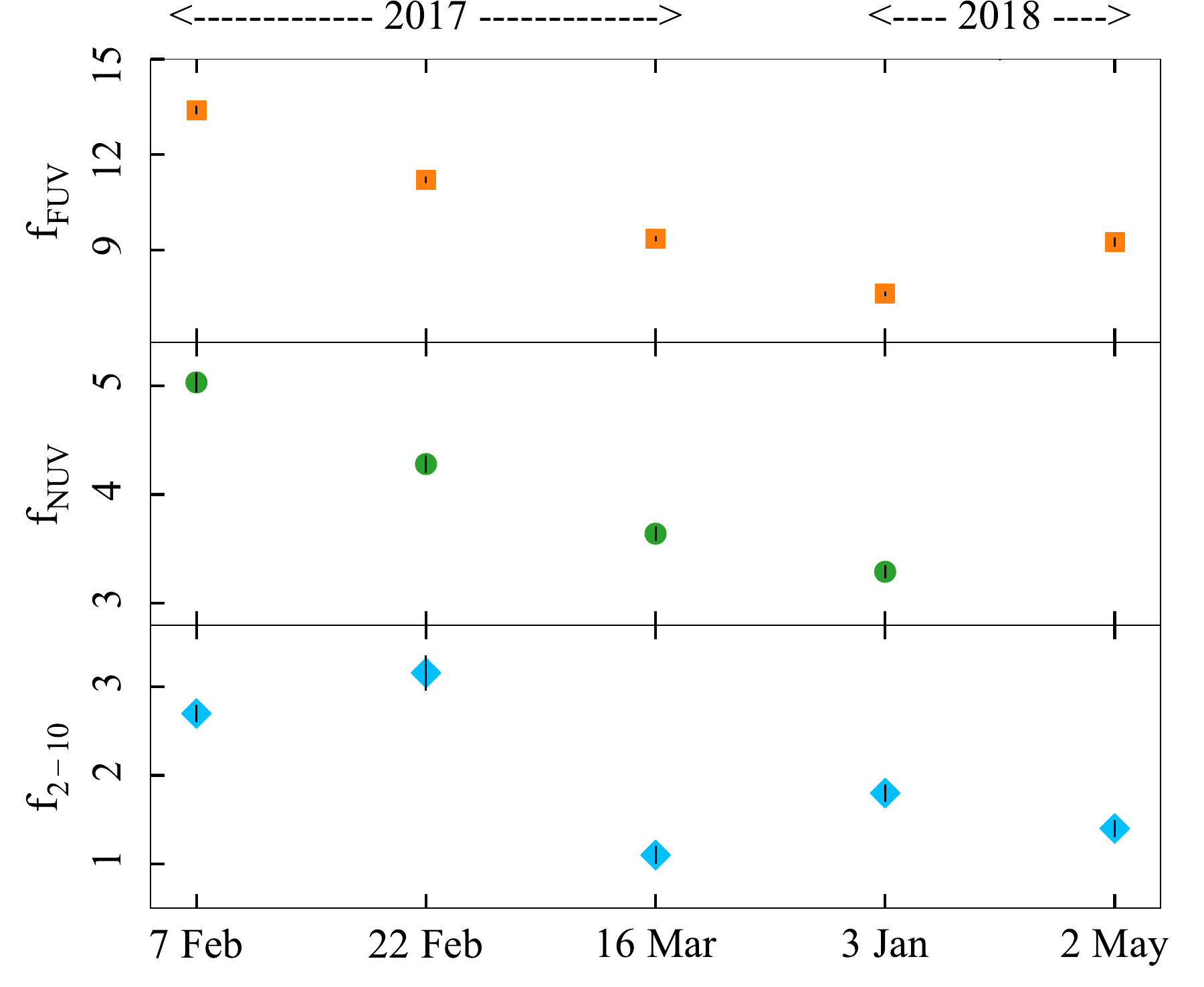}
\caption{Top: FUV-BaF$_2$ flux in the unit $10^{-14}$ \ergA. Middle: NUV-B15 filter flux in the unit $10^{-14}$ \ergA. Bottom: Unabsorbed X-ray Flux in the $2-10$ keV energy band in the unit of $10^{-10}~\rm erg~s^{-1}$. The black vertical lines are the error on the FUV/NUV fluxes.  
\label{fig:uvx_flux}}
\end{figure}

We found the X-ray photon index varying from $\sim 1.56-1.87$, heavily absorbed by the neutral absorber of column $1.2-3.4 \times 10^{23}~\rm cm^{-2}$. This high absorption leads to a significant suppression of the soft excess. We found a weak presence of soft excess, which most likely originates in photo-ionization of regions far away from the central engine and is not directly correlated with the primary continuum flux. 

The mass of the central black hole has some uncertainty. It could be between $2.5 \times 10^{6} - 5.6 \times 10^{7}  \msun$ \citep{williams2023assessing} from estimates made using different methods: circum-nuclear gas dynamics  \citep{hicks2008circumnuclear}, stellar dynamics \citep{onken2014black,roberts2021black}, $H_\beta$ reverberation mapping \citep{bentz2006reverberation,bentz2022broad,de2018velocity} and X-ray reverberation mapping \citep{zoghbi2019revisiting}. The most recent measured mass of $1.7\pm 0.4 \times 10^{7}\msun$ \citep{bentz2022broad}, gives  an  Eddington ratio varying from $0.07-0.33$. For a mass of $2.5 \times 10^{6}\msun$, the Eddington ratio will be in the range of $0.4-2.3$ and for a mass of $5.6\times 10^{7}\msun$, the Eddington ratio will be in the range of $0.02-0.1$. We find that the relative contribution of the intrinsic accretion disc emission to the bolometric emission ($L_{bol}$) is higher than that of the X-ray emission (see Fig.~\ref{fig:all_eem}). However, this bolometric luminosity is subject to some uncertainty because of the high intrinsic extinction. As mentioned earlier, we modeled the STIS spectrum, which has a continuum flux similar to the UVIT, for estimating intrinsic extinction. Although these non-simultaneous observed spectra are similar, the intrinsic shape of the spectra and the amount of obscuration could be different. At this point, we cannot rectify this systematic uncertainty.

Based on our joint UV/X-ray spectroscopy, we derived the SED of NGC~4151 at five epochs. The absorption-corrected SEDs are compared in Figure~\ref{fig:all_eem}. 
We show the observed FUV and NUV flux and the unabsorbed $2-10$~keV X-ray flux in Figure~\ref{fig:uvx_flux}. 
In the top and middle panels in Figure~\ref{fig:uvx_flux}, we show the observed flux obtained from the circular area of radius $8.3^{\prime \prime}$ using the  FUV and NUV filters at mean wavelengths of 1541\angs~ and 2197\angs, respectively. The variation in the flux in these two bands appears correlated. This is consistent with the previous observations where a strong correlation is observed within the UV/optical bands \citep{edelson2017swift}. In the bottom panel of Fig.~\ref{fig:uvx_flux}, we show the unabsorbed X-ray flux in the $2-10$ keV energy band. The flux varies between $1.3-3.2 \times 10^{-10}$ \ergS. This is near the historical peak observed during the 1993 December  ($\sim 3.6 \times 10^{-10}$\ergS; \citealt{1996ApJ...470..364E}). Therefore, during our observations, the nuclear flux could be approaching the fourth maximum since 1993. The X-ray and UV flux variations are not correlated, and the flux in the X-ray band remains within $\sim 30\%$ except for the obs3, where the flux drops by a factor of $\sim 2$. Earlier works on short-term X-ray/UV/optical variability have shown a correlation in NGC~4151 where the  X-rays are observed to lead UV emission by $\sim 3$ days \citep{edelson2017swift}. However, on timescales of months, we do not observe a correlation though the number of observations is only five in our case.

 \subsection{Variations in the intrinsic reddening}
\begin{figure}[ht!]
 \epsscale{1.1}
\plotone{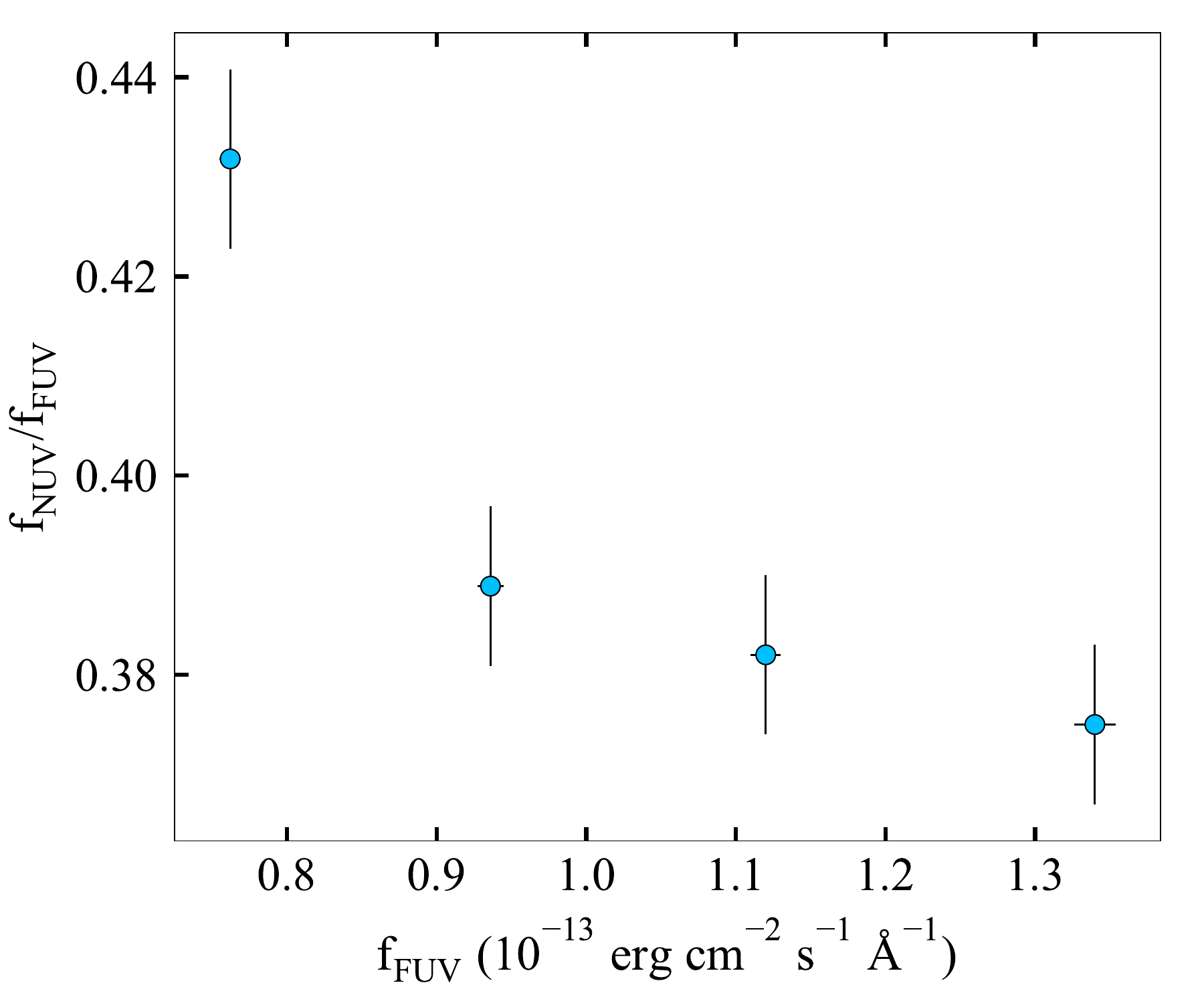}
\caption{Variation in the UV color $FUV-NUV$ with the observed FUV flux.
\label{fig:fuvflux_color}}
\end{figure}

The variation in the intrinsic UV continuum ($\sim 5-10 \times 10^{-9}$~\ergS) could be caused by our assumption of constant intrinsic reddening. In Figure~\ref{fig:fuvflux_color}, we show the ratio of observed NUV and FUV flux, a measure of $FUV-NUV$ color, as a function of observed FUV flux. The color decreases with increasing FUV flux, i.e.,  for higher FUV flux, the flux increase in the FUV band is more than that in the NUV band. Such a change can easily be caused by the variations in the intrinsic reddening as the FUV emission suffers more extinction than the NUV emission. 
Thus,  the extinction appears to be anti-correlated with the continuum flux over a year during our \asat{} observations. Using optical spectroscopic monitoring of NGC~4151 during 1996--2001, \citet{rakic2017intrinsic} found that the Balmer decrement approaches 3, as expected for a pure photoionization model with no reddening, and the Balmer decrement is also strongly anti-correlated with the continuum flux at $5100\angs$ (see their Fig.~4). These findings clearly suggest that variations in the internal reddening play a significant role in flux variability.  

%
%
The timescales of continuum flux variation over the years indicate that the variation in the UV flux we observed could be at least partly due to reddening. Since we lack information on the Balmer decrement during our observations and the quality of the UVIT data do not allow an exact estimate of the intrinsic reddening, we were unable to eliminate completely the effect of this extinction for all the observations. Therefore, we tested the possibility of varying intrinsic extinction with our UVIT data. As shown in Fig.~\ref{fig:stis_spec} (left), our FUV-grating spectrum is nearly identical to that from \textit{HST}/STIS, which we used to estimate the intrinsic extinction. Thus, the intrinsic continuum derived for obs4 is most likely a true representation of the disk emission.

\begin{figure}[ht!]
 \epsscale{1.1}
\plotone{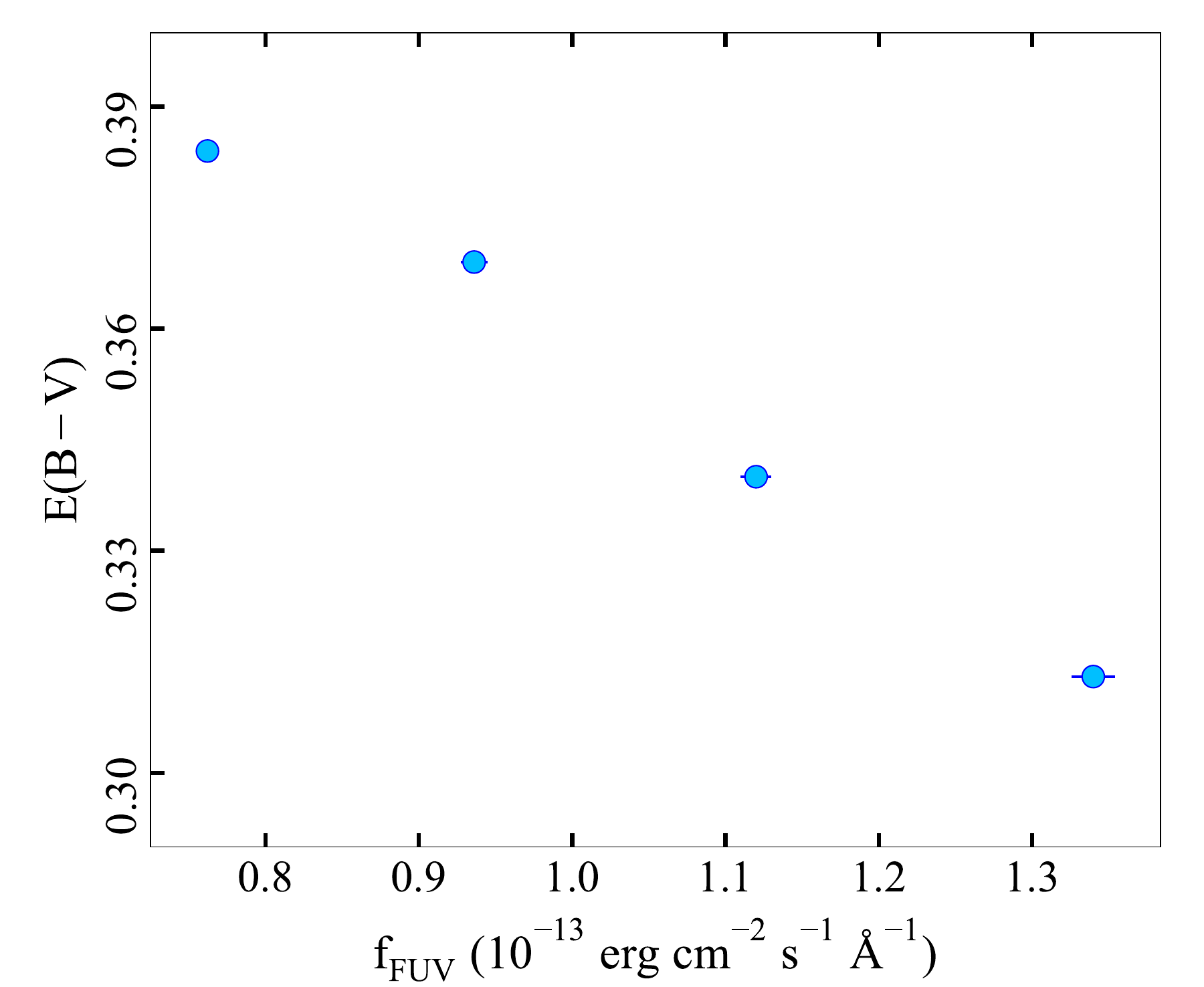}
\caption{Variation in the $E(B-V)$ with the observed FUV flux when the intrinsic continuum flux is fixed as obtained in obs4.  
\label{fig:balmdecvari_fuvflx}}
\end{figure}

To estimate the intrinsic emission for the other four observations, we assumed the disk spectrum to be constant, the same as that for obs4. We refitted the UVIT data from each observation, except obs4, by varying the intrinsic reddening only. We found that the Balmer decrement decreases with the increasing observed FUV or NUV flux, but the fit resulted in reduced $\chi^2$ more than 6. Therefore, first, we varied the line fluxes in obs5 along with the Balmer decrement. We obtained the best fit Balmer decrement of $\sim 4.84$ ($E(B-V) = 0.37$). Next, we used these updated emission line fluxes for obs1, obs2, and obs3 with a variable relative normalization. This resulted in acceptable fits with the relative line normalizations $\sim 0.98$ (obs3), 0.92 (obs2), and 0.86 (obs1). We obtained the Balmer decrement of 4.43 (obs1), 4.62 (obs2), 4.84 (obs3), and 4.83 (obs5). 
In Fig.~\ref{fig:balmdecvari_fuvflx}, we show the variation in extinction with observed FUV flux, assuming a constant disk and variable line emission. The $E(B-V)$ can be seen to anti-correlate with the observed FUV flux.

The total emission line fluxes increased by about $0.6\%$ (obs1), $8\%$ (obs2), $20\%$ (obs3 and obs5)  compared to that for obs4. On the other hand, from Fig.~\ref{fig:uvx_flux}, it can be seen that the total FUV flux, from obs1 to obs4, decreased in a uniform manner, i.e., by $\sim 20\%$, between the consecutive observations. Therefore, the line fluxes we measured in obs1 and obs2 by fixing the disk continuum are increasing disproportionately with the increase in total observed flux. Usually, during the low to medium flux state, the emission line flux is observed to correlate with the continuum flux \citep{ulrich1996month,shapovalova2008long,chen2023broad}. 
 The \ion{C}{4}  emission line is observed to be delayed with respect to the continuum emission by $\sim 3$ days \citep{ulrich1996month,metzroth2006mass}, which is less than the interval between our observations. This rules out the time delay between the continuum and the emission line as a possible cause of the non-uniform increase in emission line flux with the continuum.  Therefore, it is likely that the apparent anti-correlation observed between line and continuum flux in obs1 and obs2 is an artifact due to the lack of emission line information in those data sets. The fall in the intrinsic extinction in obs1 and obs2 is not likely as steep as estimated.   
%

Many AGN, including the changing-look AGN, show the trend of 'bluer-when-brighter' \citep{green2022time,guo2024changing}.  We examined the difference spectrum using the obs4 and obs5 grating spectra. We modeled it with intr\_ext, \dbb, and three emission lines (as these lines are variable between the observations). We fixed the Balmer decrement at 4.95. We found the disk temperature, $kT_{in}>3.4$, and normalization $ 7.3_{-4}^{+1800}\times 10^8$. As we could not constrain the temperature, it is inconclusive whether the difference spectrum is consistent with the disk.   
Thus, we conclude that the observed variation in the UV flux could be a combined effect of the change in the intrinsic shape of the continuum (`bluer-when-brighter') and internal reddening. 

\subsection{UV reddening and X-ray absorption}

We found a large neutral absorbing column ($\sim 10^{23}\rm ~cm^{-2}$) existing along the line of sight to the X-ray source. This would suggest much higher UV extinction ($\rm E(B-V) = 14  ~for~ N_H = 10^{23} ~cm^{-2}$) with the Galactic dust-to-gas ratio (Eq.~\ref{eq3:nhgal}). In our observations, the Balmer decrement of $\sim 4.95~\rm (E(B-V) = 0.38$; using Eq.~\ref{eq1:intred}) is equivalent to a column density of $\sim 2.6\times 10^{21}~\rm cm^{-2}$ in the UV following Eq.~\ref{eq3:nhgal}. Apparently, the column estimated from the UV spectrum is much smaller than that obtained from the X-ray spectrum. In the case of the two intermediate Seyferts NGC~7582 and NGC~5506, \citet{maccacaro1982x} found that the X-ray absorbing column is $\sim 10$ times larger than the $N_H$ measured using the reddening of optical continuum and Balmer lines.  This discrepancy can be attributed to different possibilities. 

The $E(B-V)$ to $N_H$ conversion relation,   Eq.~\ref{eq3:nhgal}, is obtained assuming a Galactic dust-to-gas ratio. The AGN environment could substantially differ from the Galactic dust compositions and grain sizes due to heating by the AGN \citep{maiolino2001dust,czerny2004extinction}. \citet{jaffarian2020relationship} studied the relation between the X-ray absorbing column and the $N_H$ estimated from Balmer decrement; they found that the obscuring column predicted from extinction assuming a Galactic dust-to-gas ratio is much lower than the column of the X-ray absorbing gas for most Seyfert 2. There is a large scatter between the $E(B-V)$ calculated from the Balmer decrement and $N_H$ obtained from X-ray observations considering all the AGN in their sample.

One of the possibilities to reconcile the discrepancy between the obscuring columns $N_{H}$ we estimated from our UV and X-ray observations is to assume a substantial amount of dust-free gas within the dust sublimation radius either in the form of weakly ionized winds from the accretion disk or the inner regions of the obscuring torus. 
While the dust-free region within the dust sublimation radius will absorb X-rays, it will cause very little or no UV reddening. The dusty gas outside the dust-sublimation radius will cause both X-ray absorption and UV reddening. 
This will result in effectively low column density for the UV reddening and large column density for the X-ray absorbing gas. This scenario requires a substantial amount of obscuring matter where the dust has been destroyed by the AGN heating. 

 Another possibility could be that the obscuration is caused by a two-phase medium where the dense dusty clumps are embedded in a low-density gas. This could be similar to an obscuring torus consisting of a number of compact dense clouds embedded in the low-density medium. A single or a small number of compact dense clouds along the line of sight can then cover the compact X-ray source, i.e., the hot corona, resulting in a high obscuring column. The UV emission arising from the accretion disk, which is extended, will largely be reddened by the low-density medium, thus resulting in a much lower obscuring column.
 Further, the dense clouds embedded in low-density medium responsible for the X-ray absorption are expected to be moving, which can then cause variations in the X-ray absorbing column, which has been observed.

\begin{figure}[ht!]
 \epsscale{1.2}
\plotone{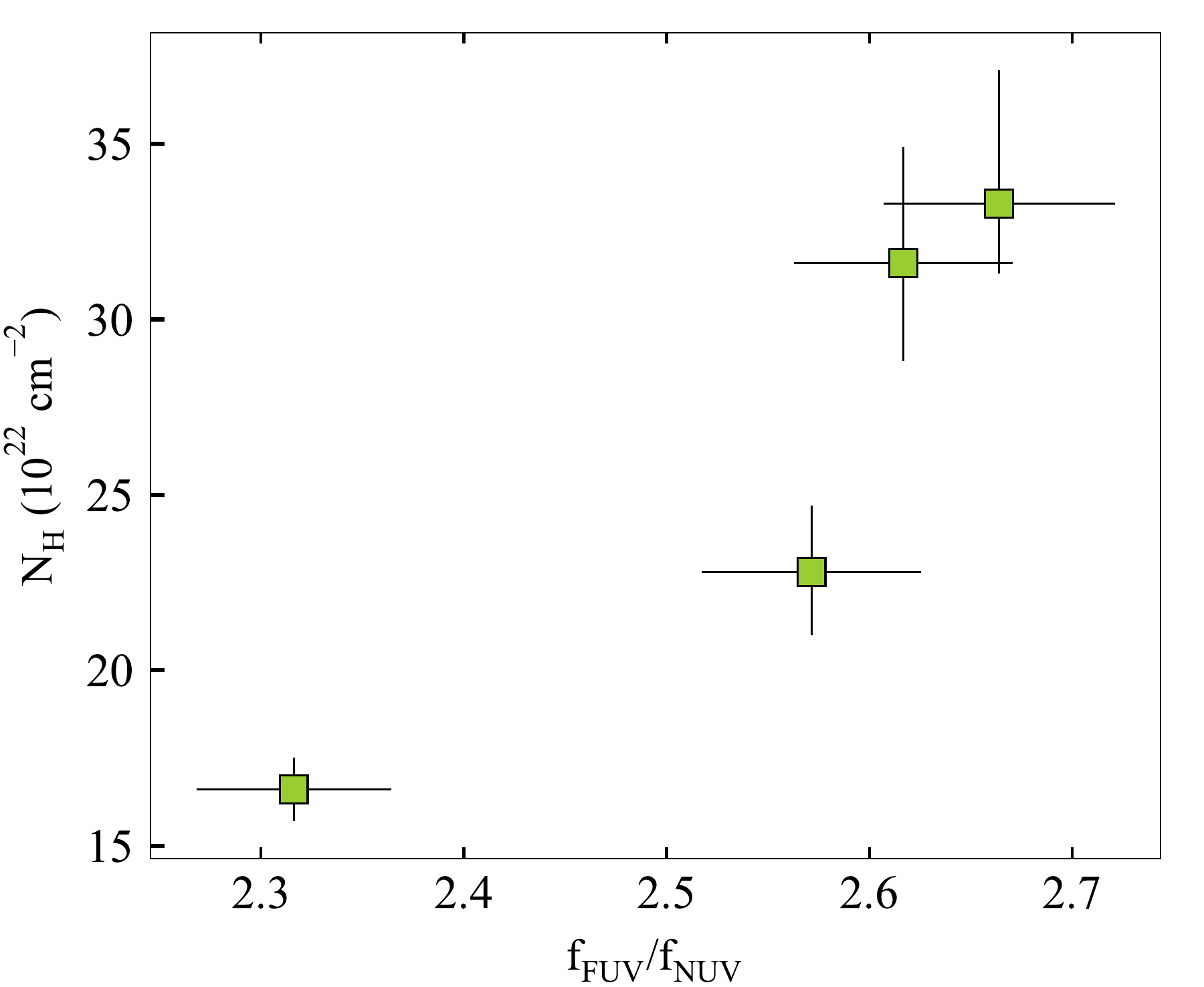}
\caption{Variation of intrinsic neutral absorbing column density with the observed $NUV - FUV$ color.  
\label{fig:color_nh}}
\end{figure}

In Figure~\ref{fig:color_nh}, we show the variations of X-ray absorbing column $N_H$ with the $NUV-FUV$ color. The apparent correlation between the X-ray $N_{H}$ and the UV color cannot be explained just by the change in the column obscuring both the X-ray source and the UV source as the increasing column will lead to redder colors contrary to the correlation. If the bluer when brighter trend seen in Fig.~\ref{fig:fuvflux_color} is at least partly caused by intrinsic variability rather than a change in the UV reddening, then it is possible that increased UV flux may be driving stronger winds which would result in the increased X-ray absorption column. This would explain the trend seen in  Fig.~\ref{fig:color_nh}.  

  \begin{figure}[ht!]
 \epsscale{1.1}
\plotone{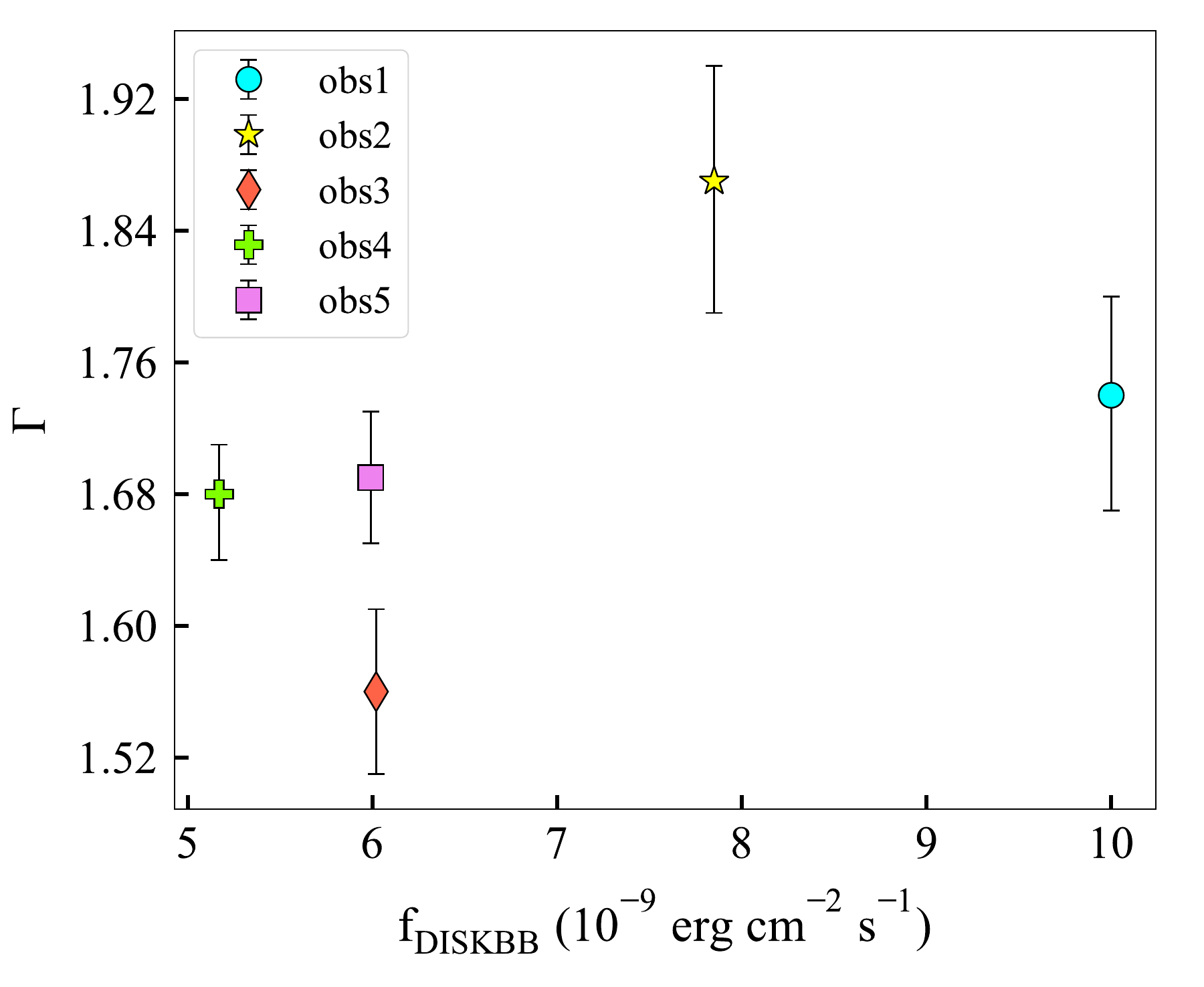}
\caption{Variation in the X-ray photon index with intrinsic disk flux (\dbb) for all five observations.  
\label{fig:uvflux_gama}}
\end{figure}

\subsection{X-ray spectral variability}
 We show the variation of the X-ray photon index with the intrinsic disk continuum emission in Figure~\ref{fig:uvflux_gama}. We observe that the total disk continuum flux increases by a factor of $\sim 2$, whereas the index remains fairly constant. 
The large time spans between the data points can wash out the effect of Compton cooling or X-ray reprocessing. Therefore, based on these observations, we could not establish any correlation on the variation in the UV/X-ray flux.


\section{conclusion}
We performed a broadband UV/X-ray spectral variability study of NGC~4151 based on five sets of \asat{} observations. The main results of our study are as follows.


\begin{enumerate}


    \item The broadband UV to X-ray spectrum of NGC~4151  primarily comprises of emission from the accretion disk  ($kT_{in} \sim 8.7-10.3$ eV), emission lines from the broad and narrow-line regions in the UV band, the primary X-ray power-law emission $\Gamma \sim 1.56-1.87$ is produced by thermal Comptonization of disk photons and a distant X-ray reflection component.  
    
    \item We obtained a high intrinsic reddening ($E(B-V) \sim 0.4$) in the UV band, generally observed in the low optical/UV flux state \citep{shapovalova2010long,rakic2017intrinsic}. 
    
    \item The $FUV-NUV$ color becomes bluer with increasing  FUV flux. This is similar to the `bluer-when-brighter' trend observed in the optical/UV spectra of many AGN and could be due to intrinsic variations in the accretion disk. Such a trend could also arise due to variations in the intrinsic reddening. We also observed increasing  X-ray absorption column with UV flux or color, this could result due to stronger winds at intrinsically bluer and higher UV emission.

    \item X-ray emission from NGC~4151 undergoes strong absorption through a large column density ($N_H\sim 1.2-3.4 \times 10^{23} ~\rm cm^{-2}$). 
    The UV emission is affected by internal reddening with  $E(B-V) \sim 0.4$, which is equivalent to an obscuring column of only $\sim 10^{21} ~\rm cm^{-2}$ assuming the Galactic dust-to-gas ratio. This $\sim 2$ orders of magnitude difference in the obscuring columns may imply that the dust-to-gas ratio of the obscurer could be very different than the Galactic value or the X-ray and the UV obscurers are not the same. We invoke two possible geometries to explain the observed discrepancy. Firstly, the obscuring medium may be divided in two zones by the dust sublimation radius. The gas within and outside the dust sublimation radius can cause X-ray absorption, while the dust outside the dust sublimation radius will be primarily responsible for the UV reddening. 
    Secondly, the obscurer could be dense, clumpy clouds embedded within low-density gas and dust. The X-ray emission could be obscured substantially by a small number of dense clouds along the line of sight to the compact hot corona, while a significant amount of the UV emission arising from an extended region may pass through the low-density regions.  

       
    \item  The X-ray fluxes during our observations are quite high ($\sim 10^{-10}$\ergS), similar to that observed during previous periods of high flux \citep{2010MNRAS.408.1851L}. We also found the presence of variable neutral and warm absorbers, usually observed in this source \citep{puccetti2007rapid,zoghbi2019revisiting}.

\end{enumerate}

\begin{acknowledgments} 
We sincerely thank an anonymous referee whose insightful comments and suggestions significantly enhanced the quality of this paper.
This work uses data from Indian
Space Science Data Centre (ISSDC) of the \textit{AstroSat} mission of the Indian Space Research Organisation (ISRO).  We acknowledge the  SXT and LAXPC POCs at TIFR (Mumbai), CZTI POC at IUCAA (Pune), and UVIT POC at IIA (Bangalore)  for providing the necessary software tools for data processing. The UVIT data were reprocessed by CCDLAB pipeline \citep{Postma_2017}. This publication used archival STIS spectrum from \textit{HST} data archive (\url{https://archive.stsci.edu/hst/search.php}). This research has used the Python and Julia packages. This research has used the SIMBAD/NED database.  S.K. acknowledges the University Grant Commission (UGC), Government of India, for financial support. Kulinder Pal Singh thanks the Indian National Science Academy for support under the INSA Senior Scientist Programme. AAZ acknowledges support from the Polish National Science Center under the grants 2019/35/B/ST9/03944, 2023/48/Q/ST9/00138, and from the Copernicus Academy under the grant CBMK/01/24.\\

\noindent{}
Facility: \asat{}, \hst{}.\\
\noindent{}
Data: The \hst/STIS spectrum presented in this article was obtained from the Mikulski Archive for Space Telescopes (MAST) at the Space Telescope Science Institute.  The specific observations analyzed can be accessed via \dataset[DOI]{http://archive.stsci.edu/doi/resolve/resolve.html?doi=10.17909/nbwv-ex98}.\\

\noindent{}
Software: CCDLAB \citep{Postma_2017}, XSPEC \citep{arnaud1996xspec}, SAOImageDS9 \citep{joye2003new}, Julia \citep{bezanson2017julia}, Astropy \citep{astropy2013astropy}.
\end{acknowledgments}

\bibliography{ngc4151}{}
\bibliographystyle{aasjournal}

\appendix

\section{STIS spectral analysis }
\label{sec:apendix}

\renewcommand{\thefigure}{A\arabic{figure}}

\setcounter{figure}{0}

We used the STIS spectrum observed on March 3, 2000, when the source was in the low flux state. We averaged the six observations of 1.2 ks exposure time each, which are nearly continuous. We used the relativistic accretion disk model \kd{} available in XSPEC for the underlying continuum. The model parameters are distance, color correction factor, black hole mass, inclination angle, inner and outer disk radius, and normalization. We fixed the mass, distance, color correction factor, inclination angle, and outer disk radius to $1.7 \times 10^{7}\msun$, $15.8$ Mpc, $2.4$, $45^\circ$, and $10^5 \rm r_g$ respectively. We used multiple Gaussian lines to account for the emission/absorption lines from the BLR/NLR region (see Table A1). Also, there are thousands of \ion{Fe}{2} emission lines that overlap with each other, making a pseudo continuum. We used \ion{Fe}{2} template model from \citet{Vestergaard_2001}, which improved the statistic by $\Delta \chi^2$ of 59 for two additional free parameters, the amplitude and the Doppler broadening. To account for the Galactic reddening, we used the XSPEC model \redd{} where we fixed the E(B-V) to 0.03, as mentioned earlier. In addition to the Galactic extinction, we tested the presence of intrinsic extinction. We used the \citet{czerny2004extinction} extinction curve with Balmer decrement as the free parameter. We obtained the best-fit value of $4.95\pm0.09$ for the Balmer decrement, and the statistic improved by $\Delta \chi^2$ of 28. As the contribution of host galaxy emission is negligible in the FUV band, we did not use this component. We obtained the best-fit inner disk radius $<4\rm~r_g$ and the mass accretion rate $2.2_{-0.2}^{+0.3} \times 10^{25}~g~s^{-1}$. The absorption lines used in the STIS spectrum are listed in Table A2. The line centroids of the absorption and emission lines are fixed during the error calculation.

 \begin{figure}[ht!]
 \epsscale{0.55}
 \plotone{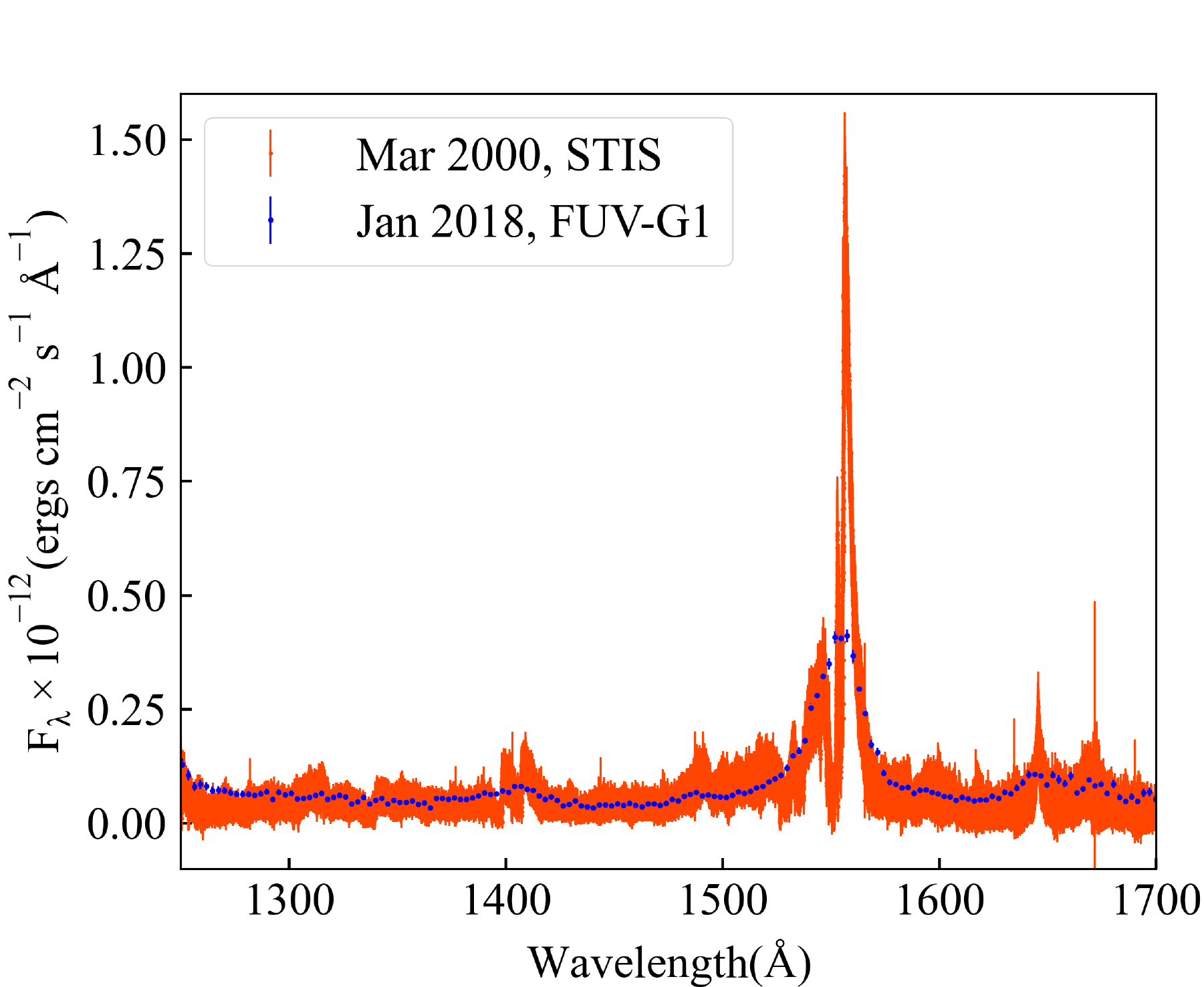}
 \epsscale{0.55}
\plotone{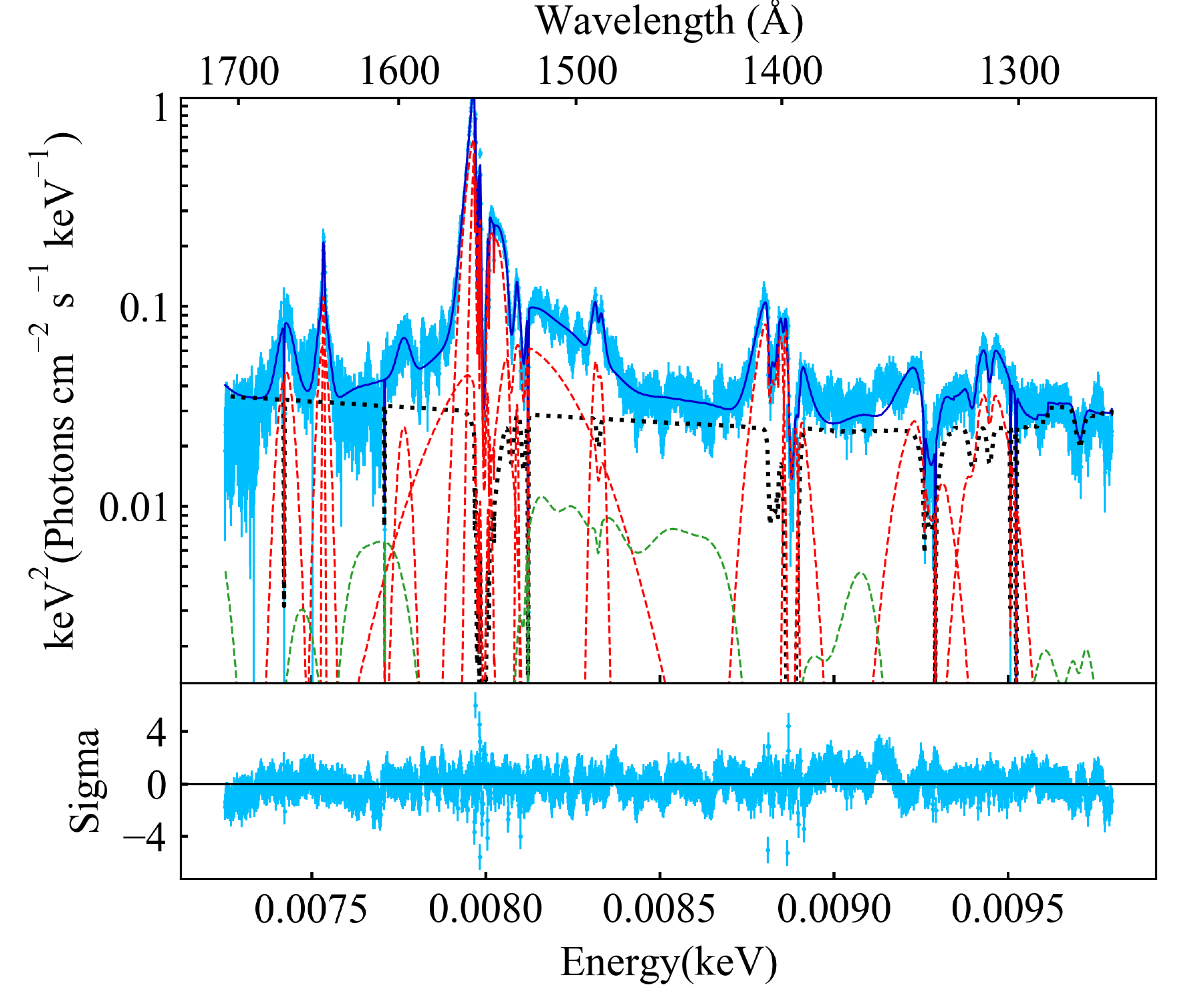}
\caption{Left: Comparison of the STIS and UVIT/FUV-G1 flux spectrum. Right: STIS spectrum: Best fit is shown in blue, reddened and absorbed continuum is shown in black dotted line, emission lines are shown in red, and \ion{Fe}{2} emission is shown in green. The (data - model)/error is shown in the bottom panel.  
\label{fig:stis_spec}}
\end{figure}

    \begin{deluxetable*}{lccccccccc}
\tablenum{A1}
\tablecaption{Best-fit parameters of the detected emission lines in STIS spectrum. $\lambda_o$ is the observed wavelength in the unit of \AA. $v_{fwhm}$ is the full-width half maxima (FWHM) of the emission line in the unit of \kms. $f_{line}$ is the line flux in the unit of $\rm photon~cm^{-2}~s^{-1}$. `(f)' -- the parameter is fixed during error calculation.}
\label{tab:stis_emis}
\tablewidth{600pt}
\tablehead{
Line & param &  &Line  & &  & &
}
\startdata
\ion{O}{1}~$\lambda 1302$\angs & & &L1~$\lambda 1549$\angs & &\\
 &$\lambda_o$ (f) & $1308$  &  & $1527$ & \\
&$v_{FWHM}$& $3702_{-207}^{+215}$ & & $386_{-126}^{+170}$\\
 &$f_{line}$& $2.4_{-0.1}^{+0.2}$ & & $0.19_{-0.08}^{+0.05}$\\
 \ion{C}{2}~$\lambda 1336$\angs & & &L2~$\lambda 1549$\angs \\
  &$\lambda_o$ (f)& $1336$ &  & $1591$ & \\
&$v_{FWHM}$& $4218$(f) & & $1637_{-272}^{+334}$ & \\
 &$f_{line}$& $2.0_{-0.1}^{+0.1}$ & & $0.6_{-0.1}^{+0.1}$\\
  \ion{Si}{4}/\ion{O}{4}]~$\lambda 1400$\angs && &\ion{He}{2}~$\lambda 1640$\angs \\
  &$\lambda_o$ (f)& $1399$  &  & $1641$ \\
&$v_{FWHM}$& $3713_{-120}^{+124}$ && $1368_{-180}^{+229}$\\
 &$f_{line}$& $6.5_{-0.3}^{+0.3}$ && $1.2_{-0.2}^{+0.2}$\\
  & $\lambda_o$ (f)& $1392$  &  & $1640$ \\
&$v_{FWHM}$& $542_{-16}^{+17}$ && $244_{-50}^{+60}$\\
 &$f_{line}$& $64_{-17}^{+23}$ && $0.4_{-0.1}^{+0.1}$\\
\ion{N}{4}]~$\lambda 1486$\angs &&& \ion{O}{3}]~$\lambda 1663$\angs \\
 &$\lambda_o$ (f)& $1485$  &  & $1664$ \\
&$v_{FWHM}$& $1123_{-110}^{+139}$ && $1987_{-226}^{+259}$\\
 &$f_{line}$& $1.1_{-0.1}^{+0.1}$ && $1.3_{-0.1}^{+0.1}$\\
\ion{C}{4}~$\lambda 1549$\angs \\
 &$\lambda_o$ (f)& $1530$  &  & \\
&$v_{FWHM}$& $13788_{-367}^{+389}$\\
 &$f_{line}$& $13.6_{-1.5}^{+0.3}$\\ 
  &$\lambda_o$ (f)& $1547$  &  & \\
&$v_{FWHM}$& $2808_{-54}^{+58}$\\
 &$f_{line}$& $44_{-5}^{+4}$\\ 
  &$\lambda_o$ (f)& $1548$  &  & \\
&$v_{FWHM}$& $1219_{-48}^{+53}$\\
 &$f_{line}$& $26_{-2}^{+2}$\\ 
\enddata
\end{deluxetable*}

\begin{deluxetable*}{lccccccccc}
\tablenum{A2}
\tablecaption{Best-fit parameters of the absorption lines detected in the STIS spectrum. `(f)' -- the parameter is fixed during error calculation.}
\label{tab:stis_abs}
\tablewidth{600pt}
\tablehead{
Line & Sigma & Strength &  Line  & Sigma & Strength\\
$10^{-3}$keV & $10^{-6}$keV & $10^{-6}$keV&$10^{-3}$keV & $10^{-6}$keV & $10^{-6}$keV
}
\startdata
$7.421$ & $0.6_{-0.4}^{+0.3}$ & $4.8_{-1.8}^{+1.3}$ & $8.815$ & $3.6_{-0.7}^{+0.7}$ & $5.9_{-1.2}^{+1.3}$\\
$7.709$ & $0.5_{-0.3}^{+0.4}$ & $3.9_{-1.8}^{+38.9}$ & $8.829$ & $10.1_{-0.9}^{+0.9}$ & $23.9_{-2.7}^{+2.8}$\\
$7.969$ & $0.8_{-0.1}^{+0.1}$ & $1.9_{-0.2}^{+0.2}$ & $8.839$ & $0.7_{-0.4}^{+0.4}$ & $0.8_{-0.5}^{+0.6}$\\
$7.976$ & $2.3_{-0.1}^{+0.1}$ & $16.4_{-0.5}^{+0.6}$ & $8.877$ & $11.5_{-0.3}^{+0.3}$ & $179_{-166}^{+192}$\\
$7.982$ & $0.8_{-0.1}^{+0.1}$ & $4.5_{-0.4}^{+0.4}$ & $9.259$ & $1.6_{-0.7}^{+1.0}$ & $2.7_{-1.4}^{+1.7}$\\
$7.994$ & $6.2_{-0.2}^{+0.2}$ & $48.7_{-1.6}^{+1.6}$ & $9.277$ & $16.3_{-1.6}^{+1.8}$ & $43.8_{-2.4}^{+5.0}$\\
$8.0089$ & $23.3_{-0.7}^{+0.7}$ & $73.9_{-3.6}^{+3.4}$ & $9.3$ & $0.38_{-0.01}^{+0.04}$  & $10$ (f)\\
$8.0091$ & $1.0_{-0.2}^{+0.2}$ & $2.7_{-0.5}^{+0.5}$ & $9.398$ & $10.2_{-2.2}^{+2.5}$ & $11.3_{-2.3}^{+2.4}$\\
$8.024$ & $0.4_{-0.2}^{+0.2}$ & $0.5_{-0.2}^{+0.2}$ & $9.447$ & $5.1_{-1.6}^{+1.9}$ & $4.5_{-1.5}^{+1.6}$\\
$8.073$ & $6.6_{-0.8}^{+0.9}$ & $9.9_{-1.3}^{+1.4}$ & $9.506$ & $0.8_{-0.5}^{+0.4}$ & $5.6_{-2.5}^{+207}$\\
$8.109$ & $7.8_{-1.3}^{+1.3}$ & $14.7_{-1.1}^{+2.3}$ & $9.523$ & $0.9_{-0.3}^{+0.4}$ & $13_{-6.79}^{+333}$\\
$8.122$ & $0.8_{-0.1}^{+0.1}$ & $8.1_{-2.3}^{-9.2}$ & $9.704$ & $10$ (f) & $10$ (f)\\
$8.324$ & $4.5_{-1.3}^{+1.6}$ & $3.6_{-1.7}^{+2.4}$\\
\enddata
\end{deluxetable*}

\end{document}